\input epsf
\documentstyle[aps]{revtex}
\tighten
\begin{document}


\preprint{HUTP-01/A063, CITA-2001-65, hep-th/0112115}

\title{Brane Gravity at Low Energy}
\author{Shinji Mukohyama}
\address{
Department of Physics, Harvard University\\
Cambridge, MA, 02138, USA
}
\author{Lev Kofman}
\address{
Canadian Institute for Theoretical Astrophysics, 
University of Toronto\\
Toronto, ON, M5S 3H8, Canada
}
\date{\today}

\maketitle

\begin{abstract} 
 Four dimensional gravity in the low energy limit of a higher
 dimensional theory has been expected to be a (generalized) Brans-Dicke
 theory. A subtle point in brane world scenarios is that the system of
 four dimensional effective gravitational equations is not closed due to
 bulk gravitational waves and bulk scalars. Nonetheless, weak gravity on
 the brane can be analyzed completely. We revisit the theory of weak
 brane gravity using gauge-invariant gravitational and scalar
 perturbations around a background warped geometry with a bulk scalar
 between two flat branes. 
 We  obtain a simple condition for the radion stabilization in
 terms of the scalar field potentials.
 We show that for general potentials of the scalar field which
 provides radion stabilization and a general conformal transformation to
 a frame in which matter on the branes are minimally coupled to the
 metric, $4$-dimensional Einstein gravity, not BD gravity, is restored
 at low  energies on either brane. 
  In contrast, in RS brane world
 scenario without a bulk scalar, low energy gravity is BD one. We
 conjecture that in general brane world scenarios with more than one
 scalar field, one will again encounter the situation that low energy 
 gravity is not described by the Einstein theory. Equipped with the
 weak gravity results, we discuss the properties of 4d brane
 gravitational equations, in particular, the value and sign of 4d
 Newton's gravitational coupling. 
\end{abstract}



\section{Introduction}


In the recent development of string/M theory~\cite{Polchinski}, branes
have been playing many important roles. The idea that our universe is a
brane in a higher dimensional spacetime has been attracting a great deal
of interest. 


Although the idea of the brane world had arisen at a phenomenological
level already in 1983~\cite{earlier-works}, it is perhaps the discovery
of the duality between M-theory and $E_8\times E_8$ heterotic
superstring theory by Horava and Witten~\cite{Horava} that made it more
attractive. It actually gives the brane world idea a theoretical
background: by compactifying six dimensions in the $11$-dimensional
theory, our $4$-dimensional universe may be realized as a hypersurface
in $5$-dimension at one of the fixed points of a $S_1/Z_2$.  
The $5$-dimensional effective theory after compactification of the six
dimensions by a Calabi-Yau manifold can be obtained, e.g. \cite{Lukas}.


Randall and Sundrum proposed two similar but distinct phenomenological
brane world scenarios~\cite{RS1,RS2}. In both scenarios the
$5$-dimensional spacetime is compactified on $S_1/Z_2$ and all matter
fields are assumed to be confined on branes at fixed points of the
$S_1/Z_2$ so that the bulk, or the spacetime region between two fixed
points, is described by pure Einstein gravity with a negative
cosmological constant. The first scenario~\cite{RS1} deals with a
three-brane with negative tension as our universe. It was shown that the
hierarchy problem may be solved in this scenario by a large redshift
(warp) factor, between our brane and another, hidden brane. In the
second scenario~\cite{RS2} a three-brane with positive tension is
considered as our universe and four-dimensional Newtonian gravity can be
realized on the brane in the limit of infinite orbifold radius. 


The distance between two branes (radion) in the phenomenological RS
scenario, as well as, more generally, moduli fields in superstring
theory, should be stabilized. For this purpose a bulk scalar field
$\Psi$ with certain potentials in the bulk and on the branes can be
introduced~\cite{GW}.  
The choice of the  bulk $U(\Psi)$ and brane potentials $V_{\pm}(\Psi)$ 
must be consistent with the 5d warp geometry \cite{Dewolfe,GKL}. In the
context of the supergravity realization of the RS scenario one has to
consider bulk scalars of the theory with an effective scalar field
potential~\cite{super1,super2}.


Thus, there is a broad class of brane world scenarios with a 5d bulk
between two end-of-the-world  branes and a bulk scalar field with bulk
and brane potentials. This is a basis for a  brane cosmology
construction in which our 4d expanding universe is located on the
visible brane.


It is expected on the general ground that dimensional reduction of a
multi-dimensional theory in the 4d low energy limit yields a
(generalized) Brans-Dicke (BD) theory. Often 4d effective theory is
derived by the integration of the action with respect to the extra
dimension, $\int dw$, as it customary in Kaluza-Klein
compactification.

In the brane world scenario one can derive effective 4d dimensional
Einstein equations on the brane, starting with the full 5d bulk
equations. Although at the qualitative level the expectation that you
will obtain an effective BD theory is justified, many quantitative
details are lost. The integration $\int dw$ must be treated with
special caution in cosmological models with brane
collisions~\cite{KKLT}.

Widespread expectation is that BD scalar is a
 manifestation of a massless bulk
scalar degree of freedom. If moduli are stabilized and no massless
scalars left in the bulk, an expectation would be that BD theory is
reduced to the pure Einstein gravity on the brane. Let us inspect a
resent progress in respect to these ideas. 

\section{Effective Einstein Equations, Weak Gravity and 4d Gravitational Coupling}

One of the most interesting results of the brane world cosmology is the
derivation of the 4d effective Einstein equations at the brane, in the
scenarios with~\cite{MW,MB} or without~\cite{BDL,SMS} a bulk scalar
field. In the models without a bulk scalar the 4d effective Einstein
equations on the brane have the structure~\cite{SMS} 
%
\begin{equation}\label{wb}
 G_{\mu\nu}=-\Lambda_4g_{\mu\nu}+8\pi \bar G_N \tau_{\mu\nu}
  +\kappa_5^4\pi_{\mu\nu}-E_{\mu\nu} \ ,
\end{equation}
where $\tau_{\mu\nu}$ comes from the brane matter,
$\Lambda_4$ is the 4d cosmological constant defined by
$\Lambda_4=\kappa_5^4\lambda^2/12-3/l_0^2$, the correction
$\pi_{\mu\nu}$ include a bilinear combination of $\tau_{\mu\nu}$, and
$E_{\mu\nu}$ is the projection of the bulk Weyl tensor. Here, $\lambda$
and $l_0$ are the brane tension and the bulk curvature scale (AdS
radius). The coupling  $\bar G_N$ which is  interpreted in \cite{SMS} as
the 4d effective Newton's gravitational constant is given by 
%
\begin{equation}\label{new1}
 \bar G_N={\kappa_5^4 \over {48 \pi}} \lambda \ .
\end{equation}
Thus gravity on a positive tension brane is described by
$4$-dimensional Einstein gravity at low energy provided the effects of
bulk gravitational waves can be neglected. On the other hand, $\bar G_N$
is negative at the negative tension brane.

In the models with a bulk scalar field (\ref{wb}) is generalized to
more complicated equations~\cite{MW,MB} 
%
\begin{equation}\label{b}
 G_{\mu\nu}=-\Lambda_4(\Psi)g_{\mu\nu}+8\pi \bar G_N(\Psi)\tau_{\mu\nu}
  +\kappa_5^4\pi_{\mu\nu}- E_{\mu\nu}+
  {{2\kappa_5^2 \over 3}}\hat T_{\mu\nu}(\Psi) \ ,
\end{equation}
where $\hat T_{\mu\nu}(\Psi)$ is the brane contribution of the scalar
field and the effective $4$-dimensional cosmological constant
$\Lambda_4$ now depends on the scalar field $\Psi$ in a non-trivial
way. Equations (\ref{b}) are not closed and must be supplemented by
the bulk equations for the scalar field with proper boundary
conditions. What is interpreted in \cite{MW,MB}
as 4d effective Newton's gravitational constant is now expressed
by the scalar field potential $V(\Psi)$ on the brane as
%
\begin{equation}\label{new2}
 \bar G_N(\Psi)={\kappa_5^4 \over {48 \pi}} V (\Psi) \ , 
\end{equation} 
which is generalization of (\ref{new1}).
One might jump to the conclusion that there is anti-gravity on the brane
with negative potential  $V<0 $~\cite{EKR}.


To address the issues of the form of the low-energy theory and the
value and  sign of $\bar G_N$ it is instructive to consider weak gravity
on the brane in a brane world scenario with flat brane but warped 5d
bulk geometry with scalar fields. We expect 4d Einstein-BD equations for
the linearized metric perturbation $h_{\mu\nu}$ around a flat background
$\eta_{\mu\nu}$ at the brane 
%
\begin{equation}\label{ebd}
 \left(\Box + \Delta K \right)  \bar{h}_{\mu\nu}=
  -16 \pi G_N \left[T_{\mu\nu} -\frac{B}{2}T\eta_{\mu\nu}
	       +\frac{2B}{3\Box}\left(\partial_{\mu}\partial_{\nu}T
	       -\frac{1}{4}\Box T\eta_{\mu\nu}\right)\right]  \ ,
\end{equation}
where $\bar{h}_{\mu\nu}$ is defined by
$\bar{h}_{\mu\nu}=h_{\mu\nu}-h\eta_{\mu\nu}/2$ with the gauge condition
$\partial^{\nu}\bar{h}_{\mu\nu}=0$.
 We have introduced a factor $B$ to
describe the Brans-Dicke type correction to the Einstein tensor gravity.
The usual BD parameter $\omega$ is introduced in the
combination $B={ 3  \over {2(3+2 \omega)}}$.
$\Delta K$ is the high-energy correction. Actually, it can be shown
that the high energy correction $\Delta K$ is consistent with the so
called higher derivative gravity, 
$\Delta K \sim \Box^2$~\cite{Mukohyama2001c}. In low energy 
approximation we neglect the high-energy correction, 
$\Delta K \approx 0$. 

Weak gravity in the Randall-Sundrum brane world (without bulk scalar)
was investigated by Garriga and Tanaka~\cite{Garriga-Tanaka} and they
showed that linearized Brans-Dicke theory is realized on the brane with 
%
\begin{equation}\label{new3}
 G_N^{\pm}=  \frac{\kappa_5^2 }{16\pi l_0} \,  
  \frac{e^{\pm d/l_0}} {\sinh(d/l_0)}   \ , \,\,\,\,\,\,  
  B_{\pm}={1 \over 2}e^{\mp 2d/l_0} \ ,
\end{equation}
where $d$ is the inter-brane distance. Einstein theory is restored only
in the limit of an infinite orbifold radius $d$, $\omega$ on the
positive tension brane becomes infinite, $B_+ =0$. 
On the other hand, formula (\ref{wb}) of Shiromizu, et.al~\cite{SMS}
shows that gravity on a positive tension brane is described by
$4$-dimensional Einstein gravity at low energies provided the effects of
bulk gravitational waves can be neglected.


In order to understand the two different-looking results
(\ref{new1}) and (\ref{new3})
in a unified manner, let us briefly
review linearized Einstein and Brans-Dicke theories around a
$4$-dimensional Minkowski background. First, because of the Poincare 
symmetry of the background it is possible to Fourier expand all
perturbations and to classify them by irreducible representations of the
little group, see Appendix for the details. 
There are three relevant classes: scalar, vector, and
tensor perturbations. Next, we can choose a suitable gauge or construct
a set of gauge-invariant variables. Third, we can write down
linearized equations for each class separately. It is easily shown that
vector perturbations should vanish. The characteristic feature of
Einstein theory is that the coupling constant between gravity and matter
is the same for scalar and tensor perturbations and 
 given by Newton's constant (see subsection 
\ref{subsec:4DEinstein}). On the other hand, in Brans-Dicke theory the
universality is spoiled by the existence of a scalar field included in
the theory since the scalar field couples with the scalar perturbations
only. The  Brans-Dicke parameter determines how the scalar
field couples to scaler perturbations. Hence the gravitational coupling
constant for scalar perturbations is a combination of Newton's constant
and the Brans-Dicke parameter while that for tensor perturbations is 
Newton's constant itself. To see this, let us set $\Delta K=0$ in
(\ref{ebd}) to obtain the linearized equation for $4$-dimensional BD
theory and split $\bar{h}_{\mu\nu}$ into tensor (transverse-traceless)
and scalar (trace) parts, 
$\bar{h}_{\mu\nu}=\bar{h}^{TT}_{\mu\nu}+ \bar{h}^{LL}_{\mu\nu} -
h\eta_{\mu\nu}/4$, with the transversal-traceless conditions upon
$h^{TT}_{\mu\nu}$. Here, $\bar{h}^{LL}_{\mu\nu}$ is given by
%
\begin{equation}
 \bar{h}^{LL}_{\mu\nu} = \frac{1}{3\Box}
  \left(\partial_{\mu}\partial_{\nu}h
   -\frac{1}{4}\Box h\eta_{\mu\nu}\right).
\end{equation}
Accordingly, equation (\ref{ebd}) is split into two parts, tensor and
scalar ones  
%
\begin{eqnarray}
 \Box \bar{h}^{TT}_{\mu\nu} & = & -16 \pi G_N  T^{TT}_{\mu\nu}  \ ,
  \nonumber\\
 \Box h & = & {16 \pi G_N}(1-2B)T \ .\label{eqn:BD-linear-decomposed}
\end{eqnarray}

Let us apply this to the issue of weak gravity on branes in the
Randall-Sundrum scenario. We can easily show by using the
$5$-dimensional Einstein equation that there are no scalar-type
perturbations in the bulk except for a zero mode called radion. 
The zero mode does not affect the value of the $4$-dimensional 
gravitational coupling constant for scalar perturbations
$G_N^{\pm}(1-2B_{\pm})$ in (\ref{eqn:BD-linear-decomposed}) since 
$\Box h=0$ for a zero mode by definition and $T=0$ for a zero mode by
the conservation equation. Hence, we can pretend as if there were no
scalar-type perturbations in the bulk, and the results (\ref{wb}),
(\ref{new1}) of Shiromizu, et.al~\cite{SMS} without ambiguities
determines the $4$-dimensional gravitational coupling constant for
scalar perturbations $\bar  G_N= G_N^{\pm}(1-2B_{\pm})= \pm {\kappa_5^2
\over {8\pi l_0}}$. 
This follows from (\ref{new3}) and from (\ref{new1}) where one has to
use flat brane tuning $\lambda=\pm{ 6 \over {\kappa^2 l_0}}$.

On the other hand, the coupling constant for tensor perturbations cannot
be determined directly since there are tensor-type perturbations in the
bulk manifested in $E_{\mu\nu}$. In other words, the system of equations
(\ref{wb}) is not closed in respect  tensor modes, and shall be
supplemented by information about bulk Weyl tensor. Actually, the
existence of tensor-type perturbations in the bulk $E_{\mu\nu}$ shifts
the $4$-dimensional gravitational coupling constant for tensor
perturbations from $\bar G_N$ as it defined in  (\ref{wb}),
(\ref{new1}),  see also \cite{SSM}.
The gravitational coupling for tensor mode is given by  $ G_N$
from (\ref{new3}). As a result, we have different $4$-dimensional
gravitational coupling constants for scalar and tensor perturbations
and, thus, the linearized Brans-Dicke theory.


Let us discuss more general case of the brane world scenario with bulk
scalar fields.

We begin with the case of a single bulk scalar field $\Psi$.
Equation (\ref{b}) describes effective 4d gravity on the brane.
We just learned that bulk gravity $E_{\mu\nu}$ is important to reproduce
correct weak gravity at the brane. Now we have to understand the effect
of bulk scalar contribution $\hat T_{\mu\nu}(\Psi)$ and
$\Lambda_4(\Psi)$ to the 4d gravity in equation (\ref{b}).

In the analysis of weak gravity one of the essential differences between
models with and without a bulk scalar field is that there are
non-trivial scalar-type perturbations in the bulk for models with a bulk
scalar field, whereas these perturbations vanish for models without a
bulk scalar field. Hence, the existence of the scalar field can shift
the $4$-dimensional gravitational coupling constant for scalar
perturbations from what is given in (\ref{new2}). On the other hand, we
already know that the existence of tensor-type perturbations in the bulk
shifts the $4$-dimensional gravitational coupling constant for tensor
perturbations. Hence, our question now is whether these shifts in the
gravitational couplings for scalar and tensor modes due to bulk gravity
$E_{\mu\nu}$ and bulk scalar $\hat T_{\mu\nu}(\Psi)$ are the
same or not. If they are the same then the effective gravitational
theory on the brane is pure Einstein theory (with a shifted Newton's
constant, which differs from (\ref{new2})). If they are not the same
then the effective theory is still Brans-Dicke theory, but still shifted
from (\ref{new2}).


The issue of weak gravity in the generalized braneworld  scenario
with a single scalar field 
was already investigated by Tanaka and Montes~\cite{Tanaka-Montes}. They
obtained very intriguing  result that 
$4$-dimensional Einstein gravity is restored at low
energy. In other words, according to the above arguments, their claim is
that shifts in the $4$-dimensional gravitational constants for scalar
and tensor perturbations bring them to the same value.


The analysis of \cite{Tanaka-Montes} can also be generalized to include
non-minimal coupling (which we denote as
$e^{-\alpha_{\pm}(\Psi_{\pm})}$)  of the brane matter to the metric in
the $5$-dimensional Einstein frame (to reflect the spirit of
dilaton-matter coupling). This non-minimal coupling was included in
the 4d Einstein equations in \cite{CGRT,MW}, where it was assumed that
matter on the brane is minimally coupled not to the metric in the 
$5$-dimensional Einstein frame but to a conformally related metric. It
is evident that the induced metric on the brane in the minimally coupled
frame should be considered as a physical metric on the brane.


Since the conformal factor depends on the bulk scalar field, metric
perturbations in the minimally coupled frame have an extra contribution
from scalar perturbations. Hence, from the above arguments about the
universality of gravitational coupling constants for scalar and tensor
perturbations in $4$-dimensional Einstein gravity it is expected that
the restoration of $4$-dimensional Einstein gravity may depend on the
choice of the minimally coupled frame. 

In the case of a single bulk scalar field we obtain
%
\begin{equation}\label{new5}
 G_N^{\pm} = \frac{\kappa_5^2 }{16\pi } \,  
  \frac{e^{-2A_{\pm}-\alpha_{\pm}}}{\int dw e^{-3A(w)}} 
  \ , \,\,\,\,\,\,  B_{\pm}=0 \  .
\end{equation}
Here, $e^{-A(w)}$ is the warp factor in the $5$-dimensional Einstein
frame, whereas the warp factor in the minimally coupled frame is
$e^{-A_{\pm}-\alpha_{\pm}/2}$. The value of $G_N^{\pm}$ is always
positive and differs from (\ref{new2}). We conclude that the coupling
with the brane energy-momentum tensor alone in the effective Einstein
equation (\ref{b}) cannot be interpreted as the 4d Newton's constant. 
The correct value of the Newton's constant is given by (\ref{new5}). 
In particular, there is no antigravity at the negative tension brane.


The  purpose of this paper, on technical side, is to perform a complete
analysis of weak gravity in the generalized braneworld world scenario in
which there is a bulk scalar field. We take into account the fact that
matter on the brane is not minimally coupled to the metric in the
$5$-dimensional Einstein frame in general and allow an arbitrary
conformal transformation to the minimally coupled frame. Despite the
above expectation, our conclusion is still the same as that of Tanaka
and Montes: $4$-dimensional Einstein gravity is restored on branes at
low energy. We formulate the necessary and sufficient condition for
brane gravity at low energy to be the Einstein theory. We also will try
to to understand the connection between the bulk scalar degrees of
freedom and the character of the brane gravity. Indeed Einstein, not BD,
gravity at the brane is restored  in the weak gravity for a
dilaton potentials in the bulk and at stabilized branes. Moreover, this
result can be extended for the second order of perturbative gravity
within a certain class of symmetric perturbations~\cite{KT}.


It is conjectured that for two or more bulk scalar fields (unless their 
potential has special properties), the weak gravity at the brane may be 
again BD gravity. We can summarize that for models without a bulk scalar
field the parameter $B$ in the low-energy gravity theory (\ref{ebd}) 
is $B=\frac{1}{2}e^{\mp d/l_0}$. In the case of a single bulk scalar field
$B=0$. For  $n > 1$ scalars in the bulk we may again expect a nonzero $B$. 
It is intriguing that  the BD parameter, which can be in principle
constrained by observations, depend on the  details of the bulk theory.

How one can understand this result? Suppose there are $n$ scalars in the
bulk. In the weak gravity limit we still expect the effective
gravitational theory of the form (\ref{ebd}). In other words, we expect 
the BD theory with a single effective BD scalar on the brane. Thus, the
effective BD scalar on the brane, in some sense, is a composite object
due to superposition of contributions from the bulk scalars. For the
case without scalars in the bulk, but the radion, we have the non-zero
value (\ref{new3}) of $B$. For the case of a single bulk scalar the
superposition of its contribution and the radion contribution to the
brane scalar gravity are precisely canceled so that $B=0$. In the case 
of more than one bulk scalars the precise cancellation is not always
expected.


The rest of the paper is rather technical and is organized as
follows. In Sec.~\ref{sec:model} we describe our model of the brane
world and define the physical energy momentum tensor of matter fields on
the branes. In the section we also specify the background around which
we analyze perturbations in Sec.~\ref{sec:perturbations}. In
Sec.~\ref{sec:comparison} we compare gravity in the brane world with
that in $4$-dimensional Einstein gravity. We show that $4$-dimensional
Einstein gravity is restored at low energies on both branes provided
there is no tachyonic mode with a boundary condition corresponding to
vanishing stress energy of matter fields on the branes. Finally,
Sec.~\ref{sec:summary} is devoted to a summary of the paper and
discussion.


\section{Model description and background}
\label{sec:model}

In this section we describe our brane world model and specify a 
background. In the next section we investigate perturbations around
this background to analyze weak gravity in the brane world model.

\begin{figure}[b]
\centering\leavevmode\epsfysize=8cm \epsfbox{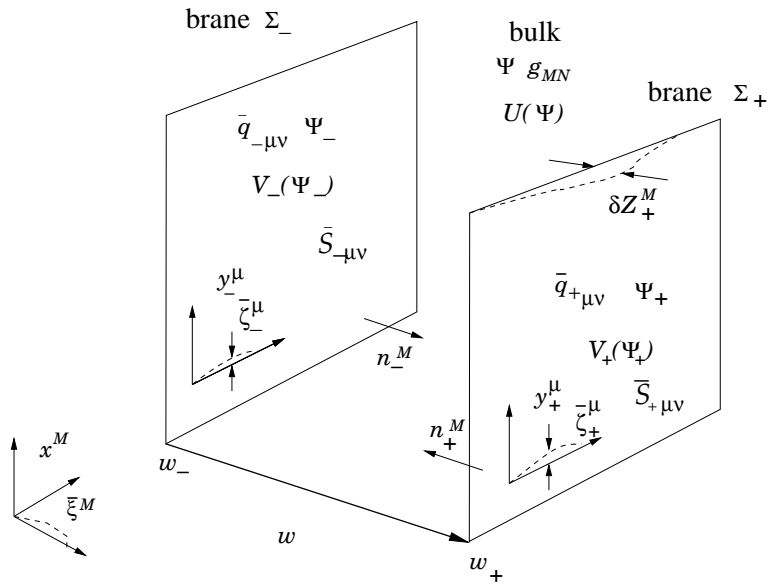}
\caption[fig1]{\label{fig1} Sketch of the brane world geometry.}
\end{figure}

Let us consider a $5$-dimensional spacetime ${\cal M}$ with
topology ${\cal M}_4\otimes S^1/Z_2$, where ${\cal M}_4$ represents 
$4$-dimensional spacetime. It is easily understood that two timelike
hypersurfaces $\Sigma_{\pm}$ corresponding to two fixed points of
$S^1/Z_2$ can be sources of gravity. In fact, because of the $Z_2$
symmetry, the first derivative of the fields can be discontinuous across
each of the fixed points and thus, the second derivative can include a
delta-function. Since Einstein equation is a second order differential
equation, inclusion of the delta-function in the second derivative of
the fields implies the existence of gravitational source at each of
fixed points. Since we would like to consider $\Sigma_{\pm}$ as a
world-volume of a brane and would like to think of one of branes as our 
$4$-dimensional universe, we assume that the $4$-dimensional intrinsic
geometry on $\Sigma_{\pm}$ is regular. On the other hand, the
$5$-dimensional geometry is not necessarily regular on
$\Sigma_{\pm}$. The $5$-dimensional region off both of $\Sigma_{\pm}$ is
usually called the bulk spacetime and we denote it by ${\cal M}_b$. 
Hence, 
%
\begin{equation}
 {\cal M} = {\cal M}_b\cup\Sigma_+\cup\Sigma_-.
\end{equation}
To describe the hypersurfaces $\Sigma_{\pm}$ we use the parametric
equations 
%
\begin{equation}
 \Sigma_{\pm}:\quad x^M = Z^M_{\pm}(y_{\pm}),
\end{equation}
where $x^M$ $(M=0,\cdots,4)$ are $5$-dimensional coordinates in 
${\cal M}$ and each $y_{\pm}$ denotes four parameters
$\{y_{\pm}^{\mu}\}$ $(\mu=0,\cdots,3)$ playing the role of the intrinsic
coordinate system on $\Sigma_{\pm}$. See Figure \ref{fig1} for
illustration.

We consider a dilaton type theory with a single scalar field to describe
gravity and assume that all other matter fields are confined to the
hypersurfaces $\Sigma_{\pm}$. Hence, in the following arguments we 
assume that the system is described by the action~\footnote{For
simplicity we do not consider the boundary of ${\cal M}_4$, but it is
easy to take it into account by imposing suitable boundary conditions
and introducing boundary terms appropriate for these boundary
conditions.} 
%
\begin{equation}
 I_{tot} = I_{EH} + I_{\Psi} + I_{matter},
  \label{eqn:total-action}
\end{equation}
where $I_{EH}$ is the Einstein-Hilbert action 
%
\begin{equation}
 I_{EH} = \frac{1}{2\kappa_5^2}\int_{\cal M} d^5x\sqrt{-g}R,
\end{equation}
$I_{\Psi}$ is the action of a scalar field $\Psi$
%
\begin{equation}
 I_{\Psi} = -\int_{{\cal M}_b} d^5x\sqrt{-g}
  \left[\frac{1}{2}g^{MN}\partial_M\Psi\partial_N\Psi
   +U(\Psi)\right]
  - \sum_{\sigma=\pm}\int_{\Sigma_{\sigma}}d^4y_{\sigma}
  \sqrt{-q_{\sigma}}V_{\sigma}(\Psi_{\sigma})
\end{equation}
and $I_{matter}$ is the action of matter fields living on the branes
%
\begin{equation}
 I_{matter} = \sum_{\sigma=\pm}\int_{\Sigma_{\sigma}}d^4y_{\sigma}
     {\cal L}_{\pm}[\bar{q}_{\pm\mu\nu},\mbox{matter}]. 
     \label{eqn:matter-action}
\end{equation}
Here, $\Psi_{\pm}$ and $q_{\pm\mu\nu}$ represent the pullback of the
scalar field $\Psi$ and the induced metric on $\Sigma_{\pm}$, and
$\bar{q}_{\pm\mu\nu}$ is the physical metric on $\Sigma_{\pm}$ specified
below. The physical metric $\bar{q}_{\pm\mu\nu}$ is actually `physical`
in the sense that because of the form (\ref{eqn:matter-action}) of the
matter action, all matter fields living in the $4$-dimensional manifolds
$\Sigma_{\pm}$ feel the geometry described by it.

The induced metric $q_{\pm\mu\nu}$ and the pullback $\Psi_{\pm}$ of
$\Psi$ on $\Sigma_{\pm}$ are defined by 
%
\begin{eqnarray}
 q_{\pm\mu\nu}(y_{\pm}) & = & 
  \left.e^M_{\pm\mu}e^N_{\pm\nu}g_{MN}\right|_{x=Z_{\pm}(y_{\pm})},
  \nonumber\\
 \Psi_{\pm}(y_{\pm}) & = & \left.\Psi\right|_{x=Z_{\pm}(y_{\pm})},
\end{eqnarray}
respectively, where 
%
\begin{equation}
 e_{\pm\mu}^M = \frac{\partial Z_{\pm}^M}{\partial y_{\pm}^{\mu}}. 
\end{equation}
We assume that the physical metric $\bar{q}_{\pm\mu\nu}$ is conformally
equivalent to the induced metric $q_{\pm\mu\nu}$: 
%
\begin{equation}
 \bar{q}_{\pm\mu\nu} = \exp[-\alpha_{\pm}(\Psi_{\pm})]q_{\pm\mu\nu}, 
\end{equation}
where $\alpha_{\pm}$ is a function of $\Psi_{\pm}$, respectively.

As shown in ref.~\cite{Mukohyama2001b}, the action above gives a
consistent variational principle including variations of the positions
of the hypersurfaces $Z_{\pm}^M$ as well as the metric $g_{MN}$, the
scalar field $\Psi$, and other matter fields. Specializing to our model
with the $Z_2$-symmetry, the equations of motion are as follows. 
%
\begin{eqnarray}
 G_{MN} & = & \kappa_5^2T_{MN}, \nonumber\\
 2K_{\pm\mu\nu} & = & 
  - \kappa_5^2\left(S_{\pm}^{\mu\nu} - \frac{1}{3}S_{\pm}q_{\pm}^{\mu\nu} 
	   \right), 
\end{eqnarray}
and 
%
\begin{eqnarray}
 \Psi^{;M}_{\ ;M}-U'(\Psi) & = & 0, \nonumber\\
 2\partial_{\perp}\Psi_{\pm} & = &
  V_{\pm}'(\Psi_{\pm}) + \frac{1}{2}\alpha_{\pm}'(\Psi_{\pm})\bar{S}_{\pm},
\end{eqnarray}
where the semicolon denotes the covariant derivative compatible with the
metric $g_{MN}$, $G_{MN}$ is the Einstein tensor for the metric
$g_{MN}$, $T_{MN}$ is the stress energy tensor of the scalar field in
the bulk, and prime is a derivative with respect to $\Psi$. 
%
\begin{equation}
 T_{MN} =  \partial_M\Psi\partial_N\Psi
	-g_{MN}\left[\frac{1}{2}g^{M'N'}\partial_{M'}\Psi\partial_{N'}\Psi
	+U(\Psi)\right],
\end{equation}
$K_{\pm\mu\nu}$ and $\partial_{\perp}\Psi_{\pm}$ are, respectively, the
extrinsic curvature of $\Sigma_{\pm}$ and the normal derivative of
$\Psi$ on $\Sigma_{\pm}$ defined by 
%
\begin{eqnarray}
 K_{\pm\mu\nu} & = &
  \frac{1}{2}\left.e^M_{\pm\mu}e^N_{\pm\nu}
        {\cal L}_{n_{\pm}}g_{MN}\right|_{x=Z_{\pm}(y_{\pm})},
  \nonumber\\
 \partial_{\perp}\Psi_{\pm} & = & 
  \left.n_{\pm}^M\partial_M\Psi\right|_{x=Z_{\pm}(y_{\pm})},
\end{eqnarray}
$n_{\pm}^M$ is the unit normal to the hypersurface $\Sigma_{\pm}$
directed towards the bulk ${\cal M}_b$, $S_{\pm\mu\nu}$ is the surface
stress energy tensor given by 
%
\begin{eqnarray}
 S_{\pm\mu\nu} & = & -V_{\pm}(\Psi_{\pm})q_{\pm\mu\nu}
  +e^{-\alpha_{\pm}(\Psi_{\pm})}\bar{S}_{\pm\mu\nu}, \nonumber\\
 \bar{S}_{\pm\mu\nu} & \equiv & 
  \bar{q}_{\pm\mu\rho}\bar{q}_{\pm\nu\sigma}\bar{S}_{\pm}^{\rho\sigma},
  \nonumber\\
 \bar{S}_{\pm}^{\mu\nu} & \equiv & \frac{2}{\sqrt{-\bar{q}_{\pm}}}
  \frac{\delta}{\delta\bar{q}_{\pm\mu\nu}}\int_{\Sigma_{\pm}}d^4y_{\pm}
  {\cal L}_{\pm}[\bar{q}_{\pm\mu\nu},\mbox{matter}],
\end{eqnarray}
and 
%
\begin{eqnarray}
 S_{\pm} & \equiv & q^{\mu\nu}S_{\pm\mu\nu} = 
  -4V_{\pm}(\Psi_{\pm}) +
  e^{-2\alpha_{\pm}(\Psi_{\pm})}\bar{S}_{\pm}, \nonumber\\
 \bar{S}_{\pm} & \equiv & \bar{q}^{\mu\nu}\bar{S}_{\pm\mu\nu}. 
\end{eqnarray}
Note that $\bar{S}_{\pm\mu\nu}$ can be called the physical surface
energy momentum tensor since it is defined in terms of the matter action
and the physical metric only. Hence, when we seek an effective
$4$-dimensional gravitational theory on the branes, we have to write the
equations in terms of $\bar{S}_{\pm\mu\nu}$ instead of $S_{\pm\mu\nu}$.

Now let us specify the background ($g_{MN}=g^{(0)}_{MN}$,
$\Psi=\Psi^{(0)}$, $Z_{\pm}^M=Z_{\pm}^{(0)M}$,
$\bar{T}_{\pm\mu\nu}=\bar{T}^{(0)}_{\pm\mu\nu}$) around which we shall
investigate perturbations in the following sections. We consider a
background with $4$-dimensional Poincare symmetry: 
%
\begin{eqnarray}
 g^{(0)}_{MN}dx^Mdx^N & = & e^{-2A(w)}(\eta_{\mu\nu}dx^{\mu}dx^{\nu}+dw^2), 
  \nonumber\\
 \Psi^{(0)} & = & \Psi^{(0)}(w), \nonumber\\
 Z_{\pm}^{(0)\mu} & = & y^{\mu}_{\pm}, \nonumber\\
 Z_{\pm}^{(0)w} & = & w_{\pm}, \nonumber\\
 \bar{S}^{(0)}_{\pm\mu\nu} & = & 0,
  \label{eqn:ansatz-background}
\end{eqnarray}
where $x^{\mu}$ ($\mu=0,\cdots,3$) represent the first four of the 
$5$-dimensional coordinates $x^M$ ($M=0,\cdots,4$) in ${\cal M}$, $w$
represents the fifth coordinate $x^4$, and $w_{\pm}$ are
constants. Here, we have redefined $V_{\pm}$ and ${\cal L}_{\pm}$ so
that $\bar{S}^{(0)}_{\pm\mu\nu}$ vanishes~\footnote{This is possible
because of the Poincare symmetry.}. Note that when we consider
perturbations around the background as in the next sections, $x^{\mu}$
on $\Sigma_{\pm}$ does not necessarily coincide with $y^{\mu}_{\pm}$.

The equations of motion for the background are as follows. 
%
\begin{eqnarray}
 3\ddot{A}+ 3\dot{A}^2 & = & \kappa_5^2\dot{\Psi}^{(0)2}, \nonumber\\
 3\ddot{A}-9\dot{A}^2 & = & 2\kappa_5^2e^{-2A}U(\Psi^{(0)})
\end{eqnarray}
and 
%
\begin{eqnarray}
 \left.e^A\dot{A}\right|_{w=w_{\pm}} & = & 
  \mp\frac{1}{6}\kappa_5^2V_{\pm}(\Psi^{(0)}_{\pm})
  \nonumber\\
 \left.e^A\dot{\Psi}^{(0)}\right|_{w=w_{\pm}} & = & 
  \mp\frac{1}{2}V'_{\pm}(\Psi^{(0)}_{\pm}),
  \label{eqn:background-junction}
\end{eqnarray}
where dots denote derivative with respect to $w$ and
$\Psi^{(0)}_{\pm}=\Psi^{(0)}(w_{\pm})$. Note that the sign in the right
hand side of (\ref{eqn:background-junction}) is due to the sign in the
following expression of the unperturbed unit normal $n^{(0)M}_{\pm}$ to 
$\Sigma^{(0)}_{\pm}$ directed towards ${\cal M}_b$. 
%
\begin{equation}
 n^{(0)M}_{\pm}\partial_M = \mp e^A\partial_w. 
  \label{eqn:unit-normal}
\end{equation}
Here, we assumed that $w_-<w_+$ and that the bulk is the region
$w_-<w<w_+$. The induced metric and the physical metric on
$\Sigma_{\pm}$ are 
%
\begin{eqnarray}
 q_{\pm\mu\nu}^{(0)} & = & e^{-2A(w_{\pm})}\eta_{\mu\nu},\nonumber\\
 \bar{q}_{\pm\mu\nu}^{(0)} & = & 
 e^{-\alpha_{\pm}(\Psi^{(0)}_{\pm})}e^{-2A(w_{\pm})}\eta_{\mu\nu}. 
\end{eqnarray}


\section{Perturbations}
\label{sec:perturbations}

In this section we investigate perturbations around the background
specified in the previous section. Namely, we consider
%
\begin{eqnarray}
 g_{MN} & = & g^{(0)}_{MN} + \delta g_{MN}, \nonumber\\
 \Psi & = & \Psi^{(0)} + \delta\Psi, \nonumber\\
 Z_{\pm}^M & = & Z_{\pm}^{(0)M} + \delta Z_{\pm}^M, \nonumber\\
 \bar{S}_{\pm\mu\nu} & = & \bar{S}^{(0)}_{\pm\mu\nu} 
 + \delta\bar{S}_{\pm\mu\nu}.
\end{eqnarray}
The unperturbed functions $\{Z_{\pm}^{(0)M}\}$ specify unperturbed
hypersurfaces $\Sigma_{\pm}^{(0)}$ and the perturbed functions
$\{Z_{\pm}^M\}$ specify perturbed hypersurfaces $\Sigma_{\pm}$. 
As pointed out in ref.~\cite{Mukohyama2000b} there are two kinds of
gauge transformations, one in the bulk and another on the branes. The
former is called $5$-gauge transformation and is of the form
%
\begin{equation}
 x^M \to x^M + \bar{\xi}^M(x). 
\end{equation}
The latter is called $4$-gauge transformation and is of the form
%
\begin{equation}
 y_{\pm}^{\mu} \to y_{\pm}^{\mu} + \bar{\zeta}_{\pm}^{\mu}(y_{\pm}).
\end{equation}

Following ref.~\cite{Mukohyama2000b}, it is straightforward to
calculate perturbations of the induced metric $q_{\pm\mu\nu}$ and the
extrinsic curvature $K_{\pm\mu\nu}$ from $\delta g_{MN}$ and
$Z_{\pm}^M$. The result for a general background is
%
\begin{eqnarray}
 q_{\pm\mu\nu} & = & q^{(0)}_{\pm\mu\nu}+\delta q_{\pm\mu\nu}, 
  \nonumber\\
 K_{\pm\mu\nu} & = & K^{(0)}_{\pm\mu\nu}+\delta K_{\pm\mu\nu}, 
        \label{eqn:perturb-q-K}
\end{eqnarray}
where 
%
\begin{eqnarray}
 \delta q_{\pm\mu\nu} & = & e^{(0)M}_{\pm\mu}e^{(0)N}_{\pm\nu}
        (\delta g_{MN} + {\cal L}_{\delta Z_{\pm}}g^{(0)}_{MN}),
        \nonumber\\
 \delta K_{\pm\mu\nu} & = &
        \frac{1}{2}n_{\pm}^{(0)M}n_{\pm}^{(0)N}
        (\delta g_{MN}+2\delta Z_{\pm M;N}) K^{(0)}_{\pm\mu\nu}        
        \nonumber\\
 & &    - \frac{1}{2}n_{\pm}^{(0)L}e^{(0)M}_{\pm\mu}e^{(0)N}_{\pm\nu}
        \left[ 2\delta\Gamma_{LMN}
        + \delta Z_{\pm L;MN} + \delta Z_{\pm L;NM}
        + (R^{(0)}_{L'MLN}+R^{(0)}_{L'NLM})\delta Z_{\pm}^{L'}\right].
\end{eqnarray}
Here, ${\cal L}$ denotes the Lie derivative defined in the 
five-dimensional spacetime, the semicolon denotes the covariant 
derivative compatible with the background metric $g^{(0)}_{MN}$, 
$R^{(0)}_{L'MLN}$ is the Riemann tensor of $g^{(0)}_{MN}$ and
$\delta\Gamma_{LMN}=
(1/2)(\delta g_{LM;N}+\delta g_{LN;M}-\delta g_{MN;L})$. 
The right hand sides of the above expressions for perturbations 
$\delta q_{\mu\nu}$ and $\delta K_{\mu\nu}$ can be evaluated on the
unperturbed hypersurface $\Sigma_{\pm}^{(0)}$ at
$x=Z_{\pm}^{(0)}(y_{\pm})$. It was shown in ref.~\cite{Mukohyama2000b} 
that $\delta q_{\pm\mu\nu}$ and $\delta K_{\pm\mu\nu}$ are invariant
under the $5$-gauge transformation and transform under the $4$-gauge
transformation as
%
\begin{eqnarray}
 \delta q_{\pm\mu\nu} & \to & \delta q_{\pm\mu\nu} 
        - \bar{\cal L}_{\bar{\zeta}_{\pm}}q^{(0)}_{\pm\mu\nu}, 
	\nonumber\\
 \delta K_{\pm\mu\nu} & \to & \delta K_{\pm\mu\nu} 
        - \bar{\cal L}_{\bar{\zeta}_{\pm}}K^{(0)}_{\pm\mu\nu}, 
\end{eqnarray}
where $\bar{\cal L}$ denotes the Lie derivative defined in the
$4$-dimensional manifold $\Sigma^{(0)}_{\pm}$. Trivially, 
$\delta S_{\pm\mu\nu}$ is also invariant under the $5$-gauge
transformation and transforms under the $4$-gauge transformation as
%
\begin{equation}
 \delta \bar{S}_{\pm\mu\nu} \to \delta \bar{S}_{\pm\mu\nu} 
        - \bar{\cal L}_{\bar{\zeta}_{\pm}}\bar{S}^{(0)}_{\pm\mu\nu}.
\end{equation}

Applying these formulas to our background, we obtain the following
expressions for $\delta q_{\pm\mu\nu}$ and $\delta K_{\pm\mu\nu}$ 
%
\begin{eqnarray}
 \delta q_{\pm\mu\nu} & = & \delta g_{\mu\nu} 
  + \partial_{\mu}\delta Z_{\pm\nu} + \partial_{\nu}\delta Z_{\pm\mu}
  - 2\dot{A}\eta_{\mu\nu}\delta Z_{\pm w}
        \nonumber\\
 \delta K_{\pm\mu\nu} & = &
  \mp\frac{1}{2}e^A\left[\delta\dot{g}_{\mu\nu}
    -\partial_{\mu}\delta g_{w\nu}-\partial_{\nu}\delta g_{w\mu} 
    +\dot{A}\eta_{\mu\nu}\delta g_{ww}\right. \nonumber\\
 & & \left.
    -2\dot{A}(\partial_{\mu}\delta Z_{\pm\nu}
    +\partial_{\mu}\delta Z_{\pm\nu})
    -2\partial_{\mu}\partial_{\nu}\delta Z_{\pm w}
    +2e^A(e^{-A})^{\cdot\cdot}\eta_{\mu\nu}\delta Z_{\pm w} \right],
\end{eqnarray}
where $\delta Z_{\pm M}=g^{(0)}_{MN}\delta Z_{\pm}^N$.

Similarly, we obtain the following expressions for perturbations of
$\Psi_{\pm}$ and $\partial_{\perp}\Psi_{\pm}$. 
%
\begin{eqnarray}
 \Psi_{\pm} & = & \Psi_{\pm}^{(0)} + \delta\Psi_{\pm},
  \nonumber\\
 \partial_{\perp}\Psi_{\pm} & = & \partial_{\perp}\Psi_{\pm}^{(0)} 
 + \delta\partial_{\perp}\Psi_{\pm}, 
\end{eqnarray}
where 
%
\begin{eqnarray}
 \delta\Psi_{\pm} & = & 
  \delta\Psi + e^{2A}\dot{\Psi}^{(0)}\delta Z_{\pm w},
  \nonumber\\
 \delta\partial_{\perp}\Psi_{\pm} & = & 
 \mp\left[e^{3A}\delta g_{ww}\dot{\Psi}^{(0)} 
     - 2e^{3A}\delta g_{ww}\dot{\Psi}^{(0)} + 2e^A\delta\dot{\Psi}
     + 2e^{2A}(e^A\dot{\Psi}^{(0)})^{\cdot}\delta Z_{\pm w}\right].
\end{eqnarray}
Again, the right hand side can be evaluated on the unperturbed
hypersurface $\Sigma_{\pm}^{(0)}$ at $x=Z_{\pm}^{(0)}(y_{\pm})$. 
It is easy to calculate perturbations of the physical metric
$\bar{q}_{\pm\mu\nu}$ on $\Sigma_{\pm\mu\nu}$ by using the following
relation
%
\begin{equation}
 \delta\bar{q}_{\pm\mu\nu}  =  
  e^{-\alpha^{(0)}_{\pm}}\delta q_{\pm\mu\nu}
  -{\alpha'_{\pm}}^{(0)}e^{-\alpha^{(0)}_{\pm}}e^{-2A_{\pm}}
  \eta_{\pm\mu\nu}\bar{q}^{(0)}_{\pm\mu\nu}\delta\Psi_{\pm},
\end{equation}
where $A_{\pm}$, $\alpha^{(0)}_{\pm}$ and ${\alpha'_{\pm}}^{(0)}$
represent $A(w_{\pm})$, $\alpha_{\pm}(\Psi_{\pm}^{(0)})$ and
$\alpha'_{\pm}(\Psi_{\pm}^{(0)})$, respectively.

We have the following $4$-gauge transformation. 
%
\begin{eqnarray}
 \delta\bar{q}_{\pm\mu\nu} & \to & \delta\bar{q}_{\pm\mu\nu}
  - \partial_{\mu}\bar{\zeta}_{\pm\nu}
  - \partial_{\nu}\bar{\zeta}_{\pm\mu},
  \nonumber\\
 \delta\tilde{K}_{\pm\mu\nu} & \to & \delta\tilde{K}_{\pm\mu\nu}
  (\mbox{invariant}), \nonumber\\
 \delta\bar{S}_{\pm\mu\nu} & \to & \delta\bar{S}_{\pm\mu\nu}
  (\mbox{invariant}),
  \label{eqn:4-gauge-tr}
\end{eqnarray}
where
$\bar{\zeta}_{\pm\mu}=\bar{q}^{(0)}_{\pm\mu\nu}\bar{\zeta}_{\pm}^{\nu}$,
%
\begin{equation}
 \delta\tilde{K}_{\pm\mu\nu} \equiv 
        \delta K_{\pm\mu\nu} - \frac{1}{2}
        (K^{(0)\rho}_{\pm\mu}\delta q_{\pm\rho\nu}
        +K^{(0)\rho}_{\pm\nu}\delta q_{\pm\rho\mu}), 
\end{equation}
and
$K^{(0)\rho}_{\pm\mu}=K^{(0)}_{\pm\mu\sigma}q^{(0)\sigma\rho}_{\pm}$. 
To see the $5$-gauge invariance of $\delta\bar{S}_{\pm\mu\nu}$ note
that we have redefined $V_{\pm}$ and ${\cal L}_{\pm}$ so that
$\bar{S}^{(0)}_{\pm\mu\nu}$ vanishes (see
eq.(\ref{eqn:ansatz-background}) and the footnote after that.). The
perturbed junction condition can be written in terms of these variables
as 
%
\begin{eqnarray}
 2\delta\tilde{K}_{\pm\mu\nu}  & = &
  -\kappa_5^2\left(\delta\tilde{S}_{\pm\mu\nu}
	-\frac{1}{3}\delta\tilde{S}_{\pm}q_{\pm\mu\nu}^{(0)}\right), 
  \nonumber\\
 \delta\tilde{S}_{\pm\mu\nu} & = & 
  -e^{-2A_{\pm}}{V'_{\pm}}^{(0)}\delta\Psi_{\pm}\eta_{\mu\nu}
  +e^{-\alpha_{\pm}}\delta\bar{S}_{\pm\mu\nu},
\end{eqnarray}
and
$\delta\tilde{S}_{\pm}=q_{\pm}^{(0)\mu\nu}\delta\tilde{S}_{\pm\mu\nu}$,
where ${V'_{\pm}}^{(0)}$ represents $V'_{\pm}(\Psi^{(0)}_{\pm})$. 
The scalar field matching condition is written as
%
\begin{equation}
 \mp\delta\partial_{\perp}\Psi_{\pm} = 
  {V''_{\pm}}^{(0)}\delta\Psi_{\pm}
  + \frac{1}{2}{\alpha'_{\pm}}^{(0)}\delta\bar{S}_{\pm},
\end{equation}
where ${V''_{\pm}}^{(0)}$ and $\delta\bar{S}_{\pm}$ represent
$V''_{\pm}(\Psi^{(0)}_{\pm})$ and
$\bar{q}^{(0)\mu\nu}_{\pm}\delta\bar{S}_{\pm\mu\nu}$, respectively.

On the other hand, the $5$-gauge transformation is
%
\begin{eqnarray}
 \delta g_{\mu\nu} & \to &  \delta g_{\mu\nu} 
  - \partial_{\mu}\bar{\xi}_{\nu} - \partial_{\nu}\bar{\xi}_{\mu}
  - 2\dot{A}\eta_{\mu\nu}\bar{\xi}_w, \nonumber\\
 \delta g_{\mu w} & \to &  \delta g_{\mu w} 
  -e^{-2A}(e^{2A}\bar{\xi}_{\mu})^{\cdot} 
  - \partial_{\mu}\bar{\xi}_w, \nonumber\\
 \delta g_{ww} & \to &  \delta g_{ww} - 2e^{-A}(e^A\bar{\xi}_w)^{\cdot}, 
  \nonumber\\
 \delta Z_{\pm M} & \to & \delta Z_{\pm M} + \bar{\xi}_M, \nonumber\\
 \delta\Psi & \to &  \delta\Psi -e^{2A}\bar{\xi}_w\dot{\Psi}^{(0)}, 
\end{eqnarray}
where $\bar{\xi}_M=g^{(0)}_{MN}\bar{\xi}^N$.

Since the background has $4$-dimensional Poincare symmetry it 
is convenient to expand the perturbations in terms of
scalar, vector and tensor harmonics in
$4$-dimensional Minkowski spacetime: 
%
\begin{eqnarray}
 \delta g_{MN}dx^Mdx^N & = & 
  ( h_{(T)}T_{(T)\mu\nu} + h_{(LT)}T_{(LT)\mu\nu}
   + h_{(LL)}T_{(LL)\mu\nu} + h_{(Y)}T_{(Y)\mu\nu})dx^{\mu}dx^{\nu}
    \nonumber\\
  & & + 2( h_{(T)w}V_{(T)\mu}+h_{(L)w}V_{(L)\mu})dx^{\mu}dw
   +h_{ww}Ydw^2, \nonumber\\
 \delta\Psi & = & \psi Y, \nonumber\\
 \delta Z_{\pm M}dx^M & = & 
 (z_{\pm(T)}V_{(T)\mu}+z_{\pm(L)}V_{(L)\mu})dx^{\mu}
 + z_{\pm w}Ydw,
 \label{eqn:harmonic-expansion}
\end{eqnarray}
and
%
\begin{equation}
 \delta\bar{S}_{\pm\mu\nu} =  
  \bar{\tau}_{\pm(T)}T_{(T)\mu\nu} + \bar{\tau}_{\pm(LT)}T_{(LT)\mu\nu} 
   + \bar{\tau}_{\pm(LL)}T_{(LL)\mu\nu} + \bar{\tau}_{\pm(Y)}T_{(Y)\mu\nu}, 
\end{equation}
where $T_{(T,LT,LL,Y)\mu\nu}$, 
$V_{(T,L)\mu}$ and $Y$ are harmonics in $4$-dimensional Minkowski
spacetime (see appendix~\ref{app:harmonics} for their definition), the
$k$-dependent Fourier coefficients $h_{(T,LT,LL,Y)}$, $h_{(T,L)w}$,
$h_{ww}$, and $\psi$ are 
functions of $w$ only, and the other coefficients $z_{\pm(T,L)}$, 
$z_{\pm w}$, and $\tau_{\pm(T,LT,LL,Y)}$ are constants. We omitted the
integration with respect to $k$. Let us expand the gauge parameters 
$\bar{\xi}^M$ and $\bar{\zeta}^{\mu}_{\pm}$, too. 
%
\begin{eqnarray}
 \bar{\xi}_Mdx^M & = & 
 (\xi_{(T)}V_{(T)\mu}+\xi_{(L)}V_{(L)\mu})dx^{\mu} + \xi_wYdw,
 \nonumber\\
 \bar{\zeta}_{\pm\mu} & = & 
 \zeta_{\pm(T)}V_{(T)\mu}+\zeta_{\pm(L)}V_{(L)\mu},
\end{eqnarray}
where $\xi_{(T,L)}$ and $\xi_w$ are functions of $w$ only and
$\zeta_{\pm(T,L)}$ are constants.

Using the harmonic expansions (\ref{eqn:harmonic-expansion}), we can
obtain the corresponding harmonic expansion of
$\delta\bar{q}_{\pm\mu\nu}$, $\tilde{K}_{\pm\mu\nu}$, $\delta\Psi_{\pm}$
and $\delta\partial_{\perp}\Psi_{\pm}$. The result is 
%
\begin{eqnarray}
 \delta\bar{q}_{\pm\mu\nu} & = & 
   \bar{\sigma}_{\pm(T)}T_{(T)\mu\nu} 
   + \bar{\sigma}_{\pm(LT)}T_{(LT)\mu\nu}
   + \bar{\sigma}_{\pm(LL)}T_{(LL)\mu\nu} 
   + \bar{\sigma}_{\pm(Y)}T_{(Y)\mu\nu}, 
    \nonumber\\
 \delta \tilde{K}_{\pm\mu\nu} & = & 
   k_{\pm(T)}T_{(T)\mu\nu} + k_{\pm(LT)}T_{(LT)\mu\nu}
   + k_{\pm(LL)}T_{(LL)\mu\nu} + k_{\pm(Y)}T_{(Y)\mu\nu},
   \nonumber\\
 \delta\Psi_{\pm} & = & \psi_{\pm}Y, \nonumber\\
 \delta\partial_{\perp}\Psi_{\pm} & = & \psi_{\pm\perp}Y, 
\end{eqnarray}
where $k$-dependent Fourier coefficients are
%
\begin{eqnarray}
 \bar{\sigma}_{\pm(T)} & = & e^{-\alpha^{(0)}_{\pm}}F_{(T)},
  \nonumber\\
 \bar{\sigma}_{\pm(LT)} & = & e^{-\alpha^{(0)}_{\pm}}\phi_{\pm(T)},
  \nonumber\\
 \bar{\sigma}_{\pm(LL)} & = & e^{-\alpha^{(0)}_{\pm}}\phi_{\pm(L)},
  \nonumber\\
 \bar{\sigma}_{\pm(Y)} & = & e^{-\alpha^{(0)}_{\pm}}
  \left[ F - 2\dot{A}\phi_{\pm w} 
   - \frac{1}{2}\eta^{\mu\nu}k_{\mu}k_{\nu}\phi_{\pm(L)}
  -{\alpha'}_{\pm}^{(0)}
  (e^{-2A}\varphi+\phi_{\pm w}\dot{\Psi}^{(0)})\right],
  \nonumber\\
 k_{\pm(T)} & = & \mp\frac{1}{2}e^{-A}(e^{2A}F_{(T)})^{\cdot},
  \nonumber\\
 k_{\pm(LT)} & = & \pm\frac{1}{2}e^AF_w,\nonumber\\
 k_{\pm(LL)} & = & \pm\frac{1}{2}e^A\phi_{\pm w},\nonumber\\
 k_{\pm(Y)} & = & \mp\frac{1}{2}
  \left\{e^{-A}(e^{2A}F)^{\cdot}+(e^A)^{\cdot}F_{ww}
   +\left[\frac{1}{2}e^A\eta^{\mu\nu}k_{\mu}k_{\nu}-2(e^A)^{\cdot\cdot}
   \right]\phi_{\pm }\right\}, \nonumber\\
 \psi_{\pm} & = & \varphi + e^{2A}\dot{\Psi}^{(0)}\phi_{\pm w},\nonumber\\
 \psi_{\pm\perp} & = & \mp\frac{1}{2}
  \left[-e^{3A}\dot{\Psi}^{(0)}F_{ww}+2e^A\dot{\varphi}
   +2e^{2A}(e^A\dot{\Psi}^{(0)})^{\cdot}\phi_{\pm w}\right].
   \label{eqn:sigma-k}
\end{eqnarray}
Here, the right hand sides of (\ref{eqn:sigma-k}) are evaluated at
$w=w_{\pm}$ respectively and have been written in terms of $5$-gauge
invariant variables defined by 
%
\begin{eqnarray}
 F_{(T)} & = & h_{(T)}, \nonumber\\
 F_w & = & h_{(T)w} - e^{-2A}(e^{2A}h_{(LT)})^{\cdot}, \nonumber\\ 
 F & = & h_{(Y)} + 2 \dot{A}X_w 
  + \frac{1}{2}\eta^{\mu\nu}k_{\mu}k_{\nu}h_{(LL)}, \nonumber\\
 F_{ww} & = & h_{ww} - 2e^{-A}(e^AX_w)^{\cdot}, \nonumber\\
 \varphi & = & \psi - e^{2A}\dot{\Psi}^{(0)}X_w,\nonumber\\
 \phi_{\pm(T)} & = & z_{\pm(T)} + \left.h_{(LT)}\right|_{w=w_{\pm}}, 
  \nonumber\\
 \phi_{\pm(L)} & = & z_{\pm(L)} + \left.h_{(LL)}\right|_{w=w_{\pm}}, 
  \nonumber\\
 \phi_{\pm w} & = & z_{\pm w} + \left.X_w\right|_{w=w_{\pm}}, 
\end{eqnarray}
where $X_w=h_{(L)w}-e^{-2A}(e^{2A}h_{(LL)})^{\cdot}$. 
It is easy to show that these variables are actually $5$-gauge
invariant, by using the following $5$-gauge transformation. 
%
\begin{eqnarray}
 h_{(T)} & \to & h_{(T)}, \nonumber\\
 h_{(LT)} & \to & h_{(LT)} - \xi_{(T)}, \nonumber\\
 h_{(LL)} & \to & h_{(LL)} - \xi_{(L)}, \nonumber\\
 h_{(Y)} & \to & h_{(Y)} +2\dot{A}\xi_w 
  + \frac{1}{2}\eta^{\mu\nu}k_{\mu}k_{\nu}\xi_{(L)}, \nonumber\\
 h_{(T)w} & \to & h_{(T)w} - e^{-2A}(e^{2A}\xi_{(T)})^{\cdot}, 
  \nonumber\\
 h_{(L)w} & \to & h_{(L)w} -\xi_w - e^{-2A}(e^{2A}\xi_{(L)})^{\cdot}, 
  \nonumber\\
 h_{ww} & \to & h_{ww} - 2e^{-A}(e^A\xi_w)^{\cdot}, \nonumber\\
 X_w & \to & X_w - \xi_w. 
\end{eqnarray}
The reason why the coefficients of $\delta\bar{q}_{\pm\mu\nu}$, 
$\delta\tilde{K}_{\pm\mu\nu}$, $\delta\Psi_{\pm}$ and
$\delta\partial_{\perp}\Psi_{\pm}$ were written by $5$-gauge-invariant
variables only is that they themselves are $5$-gauge-invariant as shown
in ref.\cite{Mukohyama2000b} for $\delta q_{\pm\mu\nu}$ and 
$\delta K_{\pm\mu\nu}$. As for the $4$-gauge transformation, while 
$\delta\tilde{K}_{\pm\mu\nu}$, $\Psi_{\pm}$ and
$\partial_{\perp}\Psi_{\pm}$ are $4$-gauge invariant,
$\delta\bar{q}_{\pm\mu\nu}$ is not. 
Its $4$-gauge transformation is 
%
\begin{eqnarray}
 \sigma_{\pm(T)} & \to & \sigma_{\pm(T)}, \nonumber\\
 \sigma_{\pm(LT)} & \to & \sigma_{\pm(LT)} - \zeta_{\pm(T)}, \nonumber\\
 \sigma_{\pm(LL)} & \to & \sigma_{\pm(LL)} - \zeta_{\pm(L)}, \nonumber\\
 \sigma_{\pm(Y)} & \to & \sigma_{\pm(Y)} 
  + \frac{1}{2}\eta^{\mu\nu}k_{\mu}k_{\nu}\zeta_{\pm(L)}. 
\end{eqnarray}
Hence, we can construct the
following doubly-gauge-invariant variables. 
%
\begin{eqnarray}
 \bar{f}_{\pm(T)} & = & \bar{\sigma}_{\pm(T)} 
  = e^{-\alpha^{(0)}_{\pm}}F_{(T)},\nonumber\\
 \bar{f}_{\pm} & = & \bar{\sigma}_{\pm(Y)} 
  + \frac{1}{2}\eta^{\mu\nu}k_{\mu}k_{\nu}\bar{\sigma}_{\pm(LL)}
  = e^{-\alpha^{(0)}_{\pm}}
  \left[ F - 2\dot{A}\phi_{\pm w} 
  -{\alpha'_{\pm}}^{(0)}
  (e^{-2A}\varphi+\phi_{\pm w}\dot{\Psi}^{(0)})\right]. 
  \label{eqn:bar-f}
\end{eqnarray}
Note that the coefficients $\bar{\tau}_{(T,LT,LL,Y)}$ of
$\delta\bar{S}_{\pm\mu\nu}$ are doubly-gauge invariant variables
by themselves.

Since there are many coefficients in the above harmonic expansions, let 
us divide these into three classes. The first class is the scalar
perturbations and consists of coefficients of $Y$, $V_{(L)\mu}$, 
$T_{(LL)\mu\nu}$ and $T_{(Y)\mu\nu}$. The second is the vector
perturbations and consists of coefficients of $V_{(T)\mu}$ and
$T_{(LT)\mu\nu}$. The last is the tensor perturbations and consists of
coefficients of $T_{(T)\mu\nu}$. It is not difficult to show that 
perturbations in different classes are decoupled from each other to
linear order. Hence, in the following we analyze perturbations in each 
class separately. Decomposition into scalar, vector and tensor modes
is commonly used in cosmology. However, usually in cosmology we use
scalar, vector and tensor representations of the isometry group
related to the symmetry of 3d space. 
Meanwhile, here we will use scalar, vector and tensor representations 
of the isometry group of 4d space-time. 

Before analyzing each class in detail, here we summarize some
differences between our formalism and one in the 
literature~\cite{Garriga-Tanaka,Tanaka-Montes}. What we have to do is
exactly the same in both formalisms: we have to solve perturbed Einstein
and scalar equations in the $5$-dimensional bulk with boundary
conditions given by Israel's junction condition and the scalar field 
matching condition on the branes. Of course, this system can be
considered as a set of inhomogeneous differential equations with $Z_2$
symmetry since the non-trivial boundary conditions can be interpreted as
source terms with the delta-function
form~\cite{Garriga-Tanaka,Tanaka-Montes,Mukohyama2001a}. Hence,
mathematically there are two approaches to this problem: one is to solve
the differential equations with the boundary conditions (or the
inhomogeneous equations) directly; another is to use the Green's
function method to solve the inhomogeneous equations by using the
complete set of homogeneous solutions.
The formalism in the literature~\cite{Garriga-Tanaka,Tanaka-Montes} is
based on the second approach. In this approach we need the complete set
of homogeneous solutions in order to construct Green's function. The set
of homogeneous solutions consists of the so called zero modes and
Kaluza-Klein modes. These homogeneous solutions satisfy the boundary
conditions corresponding to vanishing matter stress energy on the
branes. Accordingly, the homogeneous solutions have discrete
spectra. 
On the other hand, in our formalism we treat the differential equations
with the boundary conditions (or the inhomogeneous equations)
directly. By doing so, we can Fourier transform all equations with
respect to the $4$-dimensional coordinates to reduce the problem into a
set of purely $1$-dimensional problems. We also classify the Fourier
components into irreducible representations (scalar, vector and tensor
modes) of the little group of $4$-dimensional Poincare symmetry. Note
that $\eta^{\mu\nu}k_{\mu}k_{\nu}$ is continuous in our formalism.

Although $\eta^{\mu\nu}k_{\mu}k_{\nu}$ is continuous and there is no
discreteness in it, the scalar and tensor Kaluza-Klein masses may still
have important roles in our formalism. Here, a scalar (or tensor) 
Kaluza-Klein mass $m_{KK}$ is, as in the literature, such a non-zero
value of $\sqrt{-\eta^{\mu\nu}k_{\mu}k_{\nu}}$ that the scalar (or
tensor, respectively) perturbation equation have a non-trivial solution
with vanishing matter stress energy on the branes (homogeneous
solution). It is expected that $\eta^{\mu\nu}k_{\mu}k_{\nu}=-m_{KK}^2$
corresponds to a resonance point in the continuous spectrum of
$\eta^{\mu\nu}k_{\mu}k_{\nu}$. Hence, there may be interesting phenomena
near the point corresponding to the Kaluza-Klein mass,
$\eta^{\mu\nu}k_{\mu}k_{\nu}\simeq -m_{KK}^2$. However, we shall not see
this in this paper since we shall adopt the expansion by a parameter
$\mu=l^2\eta^{\mu\nu}k_{\mu}k_{\nu}$, where $l$ is a length scale of the
model. Further analysis near the resonance may be an interesting future
work.

\subsection{Scalar perturbations}
\label{subsec:scalar}

For scalar perturbations, we can use the the background equation to
reduce the perturbed junction condition to 
%
\begin{eqnarray}
 2\bar{\tau}_{\pm(Y)} & = & 
  3\eta^{\mu\nu}k_{\mu}k_{\nu}\bar{\tau}_{\pm(LL)}, \nonumber\\
 \phi_{\pm w} & = & 
  \mp\kappa_5^2e^{-A-\alpha^{(0)}_{\pm}}\bar{\tau}_{\pm(LL)}. 
  \label{eqn:scalar-junction}
\end{eqnarray}
Here, we have assumed that $k^{\mu}\ne 0$ since a mode with
$k^{\mu}=0$ preserves $4$-dimensional Poincare symmetry and can be
included in the background. The first equation is a gauge-invariant 
expression of the perturbed conservation equation. The second equation
gives the gauge-invariant perturbation $\phi_{\pm w}$ of the position of
each brane in terms of matter perturbations on the brane. The perturbed
matching condition of the scalar field at $w=w_{\pm}$, respectively, is
%
\begin{equation}
 \mp\left[-e^{3A}\dot{\Psi}^{(0)}F_{ww}+2e^A\dot{\varphi}
   +2e^{2A}(e^A\dot{\Psi}^{(0)})^{\cdot}\phi_{\pm w}\right]
 = {V''_{\pm}}^{(0)}(\varphi+e^{2A}\dot{\Psi}^{(0)}\phi_{\pm w})
 + 2{\alpha'_{\pm}}^{(0)}e^{2A}\bar{\tau}_{\pm(Y)},
 \label{eqn:scalar-matching}
\end{equation}
where, as we already stated, ${V''_{\pm}}^{(0)}$ and
${\alpha'_{\pm}}^{(0)}$ are abbreviations of 
$V''_{\pm}(\Psi^{(0)}_{\pm})$ and $\alpha'_{\pm}(\Psi^{(0)}_{\pm})$,
respectively.

For scalar perturbations, the perturbed Einstein equations in the bulk 
consist of the following set of equations for $5$-gauge invariant
variables. 
%
\begin{eqnarray}
 F_{ww} & = & -2F, \nonumber\\
 \varphi & = & -\frac{3e^{2A}}{2\kappa_5^2\dot{\Psi}^{(0)}}\dot{F}, 
\end{eqnarray}
and
%
\begin{equation}
 \ddot{F} 
  + \left[2\dot{A}
     -\frac{(e^A)^{\cdot\cdot\cdot}}{(e^A)^{\cdot\cdot}}\right]\dot{F}
  - \left[2e^{-A}(e^A)^{\cdot\cdot}+\eta^{\mu\nu}k_{\mu}k_{\nu}\right]F 
  = 0. \label{eqn:scalar-bulkeq}
\end{equation}

In the following arguments, we shall consider modes with
$\eta^{\mu\nu}k_{\mu}k_{\nu}=0$. These modes represent massless fields
on branes and are called zero modes. As we shall see in the next
section, for scalar perturbations it is actually sufficient to
analyze zero modes in order to compare the low energy behavior of
gravity in the brane world scenario with $4$-dimensional Einstein
gravity.

Setting $\eta^{\mu\nu}k_{\mu}k_{\nu}=0$, it is easy to find a zero-mode
solution $F=F^{(1)}\equiv(e^A)^{\cdot}$ of
(\ref{eqn:scalar-bulkeq}). Another independent solution $F=F^{(2)}$ can
be easily found by using the fact that the Wronskian 
$\Delta\equiv\dot{F}^{(1)}F^{(2)}-F^{(1)}\dot{F}^{(2)}$ obeys the 
equation $\dot{\Delta}=-B_1\Delta$, where $B_1$ is the coefficient of 
$\dot{F}$ in (\ref{eqn:scalar-bulkeq}). The solution to the Wronskian
equation is $\Delta\propto e^{-2A}(e^A)^{\cdot\cdot}$, and thus 
%
\[
 F^{(2)} \propto (e^A)^{\cdot}
 \int dw\frac{\ddot{A}+\dot{A}^2}{\dot{A}^2}e^{-3A}. 
\]
Hence, a general solution is 
%
\begin{equation}
 F(w) = C_1(e^{A(w)})^{\cdot}
 + C_2(e^{A(w)})^{\cdot}
 \int_0^wdw'\frac{\ddot{A}(w')+\dot{A}(w')^2}{\dot{A}(w')^2}e^{-3A(w')}. 
 \label{eqn:sol-scalar}
\end{equation}

For the solution (\ref{eqn:sol-scalar}), the perturbed matching
condition (\ref{eqn:scalar-matching}) of the scalar field is 
%
\begin{equation}
 \left(C_1 + C_2B(w_{\pm}) \pm 2\kappa_5^2e^{-2A_{\pm}-\alpha^{(0)}_{\pm}}
  \bar{\tau}_{\pm(LL)}\right)\tilde{B}_{\pm} = 0,
 \label{eqn:C1-C2-bc}
\end{equation}
where
%
\begin{eqnarray}
 B(w) & = &
 \int_0^wdw'\frac{\ddot{A}(w')+\dot{A}(w')^2}{\dot{A}(w')^2}e^{-3A(w')}
 + \frac{e^{-3A(w)}}{\dot{A}(w)}, \nonumber\\
 \tilde{B}_{\pm} & = & 
  \left.(\ddot{A}+\dot{A}^2)
   \left(\ddot{\Psi}^{(0)}+\dot{A}\dot{\Psi}
   \pm\frac{1}{2}e^{-A}{V''_{\pm}}^{(0)}\dot{\Psi}\right)\right|_{w=w_{\pm}}
   \nonumber\\
  & = & \frac{\kappa_5^2}{12}e^{-4A}V^{'(0)2}_{\pm}
  \left({U'_{\pm}}^{(0)}
     +\frac{\kappa_5^2}{3}V_{\pm}^{(0)}{V'_{\pm}}^{(0)}
		 -\frac{1}{4}{V'_{\pm}}^{(0)}{V''_{\pm}}^{(0)}\right),
  \label{eqn:BandtildeB}
\end{eqnarray}
and $A_{\pm}$, $\alpha^{(0)}_{\pm}$ and ${U'_{\pm}}^{(0)}$ are
abbreviations of $A(w_{\pm})$, $\alpha_{\pm}(\Psi^{(0)}_{\pm})$ and
$U'(\Psi_{\pm}^{(0)})$, respectively. Here we have used the perturbed
junction condition (\ref{eqn:scalar-junction}) and the background 
equations to simplify the equation. Hence, if $\tilde{B}_{\pm}\ne 0$
then the coefficients $C_1$ and $C_2$ are uniquely determined as
%
\begin{eqnarray}
 C_1 & = & \frac{2\kappa_5^2}{B(w_+)-B(w_-)}
  \left[B(w_-)e^{-2A_+-\alpha^{(0)}_+}\bar{\tau}_{+(LL)}+
   B(w_+)e^{-2A_--\alpha^{(0)}_-}\bar{\tau}_{-(LL)}\right], \nonumber\\
 C_2 & = & -\frac{2\kappa_5^2}{B(w_+)-B(w_-)}
  \left[e^{-2A_+-\alpha^{(0)}_+}\bar{\tau}_{+(LL)}+
   e^{-2A_--\alpha^{(0)}_-}\bar{\tau}_{-(LL)}\right].
\end{eqnarray}
Note that there are no further conditions and that this solution
represents a set of gauge invariant perturbations. 

For further analysis
it is rather important that  $\tilde{B}_{\pm}\ne 0$. 
The rare case of $\tilde{B}_{\pm}=0$ would correspond to the situation 
when scalar brane matter perturbations are not coupled with the bulk
perturbations. It takes place, for instance for the RS models with
constant $U$ and $V_{\pm}$. 
In general, the forms of the potentials not necessarily lead to
$\tilde{B}_{\pm}=0$. 
It is interesting to check that $\tilde{B}_{\pm}=0$ in the 5d reduction
of  HW theory \cite{Lukas} with  
$U(\Psi)=\beta {1 \over 6} \exp{( -2 \sqrt{2}\kappa_5 \Psi)}$ 
and $V_{\pm}=\pm \beta \sqrt{2} \exp{( - \sqrt{2}\kappa_5 \Psi)}$,
$\beta$ is a constant. For these potentials the radion is not stabilized
in the sense that the inter-brane distance is arbitrary.

Therefore, by using the formula (\ref{eqn:bar-f}), the doubly gauge
invariant variable $\bar{f}$ representing perturbations of the physical
metric is written as 
%
\begin{equation}
 \bar{f}_{\pm} = - \frac{\kappa_5^2e^{-2A_{\pm}-\alpha^{(0)}_{\pm}}}
  {\int^{w_+}_{w_-}dw'e^{-3A(w')}}
  \left[e^{-2A_+-\alpha^{(0)}_+}\bar{\tau}_{+(LL)}+
   e^{-2A_--\alpha^{(0)}_-}\bar{\tau}_{-(LL)}\right].
  \label{eqn:barf-result-k2=0}
\end{equation}
Here, we have used the fact that $\dot{B}=-2e^{-3A}$ to simplify
$B(w_+)-B(w_-)$, and we have also used the background equation.

Although we worked in a sector with $\eta^{\mu\nu}k_{\mu}k_{\nu}=0$, the
result (\ref{eqn:barf-result-k2=0}) can be considered as the lowest
order term in the expansion of $\bar{f}_{\pm}$ by the parameter
$\mu\equiv l^2\eta^{\mu\nu}k_{\mu}k_{\nu}$, where $l$ is a
characteristic length scale of the model. (See (\ref{eqn:expansion-mu})
for the expansion of tensor perturbations.) Hence, the above result for
$\mu=0$ is enough to give the following expression of the lowest order
term in the $\mu$ expansion for a small non-zero $\mu$. 
%
\begin{equation}
 \bar{f}_{\pm} = - \frac{\kappa_5^2e^{-2A_{\pm}-\alpha^{(0)}_{\pm}}}
  {\int^{w_+}_{w_-}dw'e^{-3A(w')}}
  \left[e^{-2A_+-\alpha^{(0)}_+}\bar{\tau}_{+(LL)}+
   e^{-2A_--\alpha^{(0)}_-}\bar{\tau}_{-(LL)}\right]
  + O(\mu). 
\end{equation}

\subsection{Vector perturbations}

For vector perturbations, the perturbed junction condition at
$w=w_{\pm}$, respectively, is
%
\begin{equation}
 F_w = 
  \mp\kappa_5^2e^{-A_{\pm}-\alpha^{(0)}_{\pm}}\bar{\tau}_{\pm(LT)},
\end{equation}
and there is no matching condition for the scalar field.

The perturbed Einstein equation in the bulk for vector perturbations 
is 
%
\begin{eqnarray}
 \eta^{\mu\nu}k_{\mu}k_{\nu}F_w & = & 0, \nonumber\\
 (e^{-A}F_w)^{\cdot} & = 0. 
\end{eqnarray}
Hence, the solution is 
%
\begin{equation}
 F_w = \left\{\begin{array}{rl}
 Ce^A, & \mbox{for $\eta^{\mu\nu}k_{\mu}k_{\nu}= 0$} \\
 0, & \mbox{for $\eta^{\mu\nu}k_{\mu}k_{\nu}\ne 0$}
 \end{array}\right. \label{eqn:vector-solution}
\end{equation}
where $C$ is a constant. For $\eta^{\mu\nu}k_{\mu}k_{\nu}=0$, the
junction condition overdetermines the coefficient $C$:
%
\begin{equation}
 C = -\kappa_5^2e^{-2A_+-\alpha^{(0)}_+}\bar{\tau}_{+(LT)}
  = \kappa_5^2e^{-2A_--\alpha^{(0)}_-}\bar{\tau}_{-(LT)}. 
\end{equation}
Hence, for $\eta^{\mu\nu}k_{\mu}k_{\nu}= 0$ we have the following
relation between $\bar{\tau}_{+(LT)}$ and $\bar{\tau}_{-(LT)}$. 
%
\begin{equation}
 e^{-2A_+-\alpha^{(0)}_+}\bar{\tau}_{+(LT)}
  + e^{-2A_--\alpha^{(0)}_-}\bar{\tau}_{-(LT)} = 0.
  \label{eqn:relation-vector}
\end{equation}
As we shall see in the next section, this relation is important for 
restoring $4$-dimensional Einstein gravity at low energies. For
$\eta^{\mu\nu}k_{\mu}k_{\nu}\ne 0$, the junction condition becomes 
%
\begin{equation}
 \bar{\tau}_{+(LT)}=\bar{\tau}_{-(LT)} = 0. 
\end{equation}

Note that for vector perturbations there is no doubly gauge invariant
perturbation of physical metric on the branes.

\subsection{Tensor perturbations}
\label{subsec:tensor}

For tensor perturbations, the perturbed junction condition at
$w=w_{\pm}$, respectively, is reduced to 
%
\begin{equation}
 (e^{2A}F_{(T)})^{\cdot} = \pm\kappa_5^2e^{A_{\pm}-\alpha^{(0)}_{\pm}}
  \bar{\tau}_{\pm(T)}
  \label{eqn:junction-tensor}
\end{equation}
and there is no matching condition for the scalar field.

The perturbed Einstein equation for tensor perturbations in the bulk 
is 
%
\begin{equation}
 e^A\left[ e^{-3A}(e^{2A}F_{(T)})^{\cdot}\right]^{\cdot}
  -\eta^{\mu\nu}k_{\mu}k_{\nu}F_{(T)} = 0. 
  \label{eqn:bulkeq-tensor}
\end{equation}

It can be shown that $\eta^{\mu\nu}k_{\mu}k_{\nu}\leq 0$ for any
solutions of the equation (\ref{eqn:bulkeq-tensor}) insofar as the
boundary condition (\ref{eqn:junction-tensor}) with
$\bar{\tau}_{\pm(T)}=0$ is satisfied~\cite{Mukohyama2001c}. The proof is
easy: we can show that 
%
\begin{equation}
 \eta^{\mu\nu}k_{\mu}k_{\nu}\int_{w_-}^{w_+}dwe^AF_{(T)}^2
  = -\int_{w_-}^{w_+}dw
  e^{-3A}\left[(e^{2A}F_{(T)})^{\cdot}\right]^2 \leq 0.
\end{equation}

In the following arguments, we shall first consider zero modes, or modes
with $\eta^{\mu\nu}k_{\mu}k_{\nu}=0$. Contrary to the scalar
perturbations, we will see in the next section that for tensor
perturbations it is not sufficient to analyze zero modes in order to 
compare the low energy behavior of gravity in the brane world scenario
with 
$4$-dimensional Einstein gravity. Hence, after analyzing zero modes, we
shall investigate non-zero modes, or so called Kaluza-Klein modes by
perturbing with respect to $\eta^{\mu\nu}k_{\mu}k_{\nu}$ around the zero
modes.

For $\eta^{\mu\nu}k_{\mu}k_{\nu}=0$, the solution of the bulk Einstein
equation (\ref{eqn:bulkeq-tensor}) is 
%
\begin{equation}
 F_{(T)}(w) = D_1e^{-2A(w)}\int_0^wdw'e^{3A(w')} + D_2e^{-2A(w)},
\end{equation}
where $D_1$ and $D_2$ are constants. By using the formula
(\ref{eqn:bar-f}), the doubly gauge invariant variable
$\bar{f}_{\pm(T)}$ representing perturbations of the physical metric can
be written as 
%
\begin{equation}
 \bar{f}_{\pm(T)} = 
  D_1e^{-2A_{\pm}-\alpha^{(0)}_{\pm}}\int_0^wdw'e^{3A(w')}
  + D_2e^{-2A_{\pm}-\alpha^{(0)}_{\pm}}. 
  \label{eqn:fT-zeromode}
\end{equation}
The junction condition overdetermines the coefficient $D_1$:
%
\begin{equation}
 D_1 = \kappa_5^2e^{-2A_+-\alpha^{(0)}_+}\bar{\tau}_{+(T)}
  = -\kappa_5^2e^{-2A_--\alpha^{(0)}_-}\bar{\tau}_{-(T)}. 
\end{equation}
Hence, for $\eta^{\mu\nu}k_{\mu}k_{\nu}= 0$ we have the following
relation between $\bar{\tau}_{+(T)}$ and $\bar{\tau}_{-(T)}$. 
%
\begin{equation}
 e^{-2A_+-\alpha^{(0)}_+}\bar{\tau}_{+(T)}
  + e^{-2A_--\alpha^{(0)}_-}\bar{\tau}_{-(T)} = 0.
  \label{eqn:relation-tensor}
\end{equation}
As we shall see in the next section, this relation is important for 
restoring $4$-dimensional Einstein gravity at low energies.

For $\eta^{\mu\nu}k_{\mu}k_{\nu}\ne 0$ it is not trivial to solve
(\ref{eqn:bulkeq-tensor}) without specifying the model
potentials. Hence, we shall perform perturbation with respect to
$\mu=l^2\eta^{\mu\nu}k_{\mu}k_{\nu}$ around the zero mode solution,
where $l$ is a characteristic length scale of the
model~\footnote{The characteristic length $l$ is determined by the
background solution~\cite{Mukohyama2001c} and determines the energy
scale below which the low energy description is valid.}. By
expanding $F_{(T)}$ and $\bar{\tau}_{(T)}$ as
%
\begin{eqnarray}
 F_{(T)} & = & F_{(T)}^{[0]}+\mu F_{(T)}^{[1]} + O(\mu^2), 
  \nonumber\\
 \bar{\tau}_{\pm(T)} & = & \bar{\tau}_{\pm(T)}^{[0]}
  +\mu \bar{\tau}_{\pm(T)}^{[1]} + O(\mu^2),
  \label{eqn:expansion-mu}
\end{eqnarray}
we shall solve the Einstein equation (\ref{eqn:bulkeq-tensor}) in the
bulk and the junction condition (\ref{eqn:junction-tensor}) order by
order. The zeroth order analysis is exactly the same as the previous
analysis of the zero modes. Hence, the zeroth order Einstein equation in
the bulk has the solution
%
\begin{equation}
 F_{(T)}^{[0]}(w) = \tilde{D}_1e^{-2A(w)}\int_0^wdw'e^{3A(w')}
  + \tilde{D}_2e^{-2A(w)},
\end{equation}
where $\tilde{D}_1$ and $\tilde{D}_2$ are constants, and the zeroth
order junction condition overdetermines the coefficient $\tilde{D}_1$: 
%
\begin{equation}
 \tilde{D}_1 = 
  \kappa_5^2e^{-2A_+-\alpha^{(0)}_+}\bar{\tau}_{+(T)}^{[0]}
  = -\kappa_5^2e^{-2A_--\alpha^{(0)}_-}\bar{\tau}_{-(T)}^{[0]},
  \label{eqn:C1-tauT0}
\end{equation}
and thus we have the following relation between
$\bar{\tau}_{+(T)}^{[0]}$ and $\bar{\tau}_{-(T)}^{[0]}$. 
%
\begin{equation}
 e^{-2A_+-\alpha^{(0)}_+}\bar{\tau}_{+(T)}^{[0]}
  + e^{-2A_--\alpha^{(0)}_-}\bar{\tau}_{-(T)}^{[0]} = 0.
  \label{eqn:relation-tauT}
\end{equation}
Note that the zeroth order analysis leaves the coefficient $\tilde{D}_2$
undetermined.

To first order in $\mu$, the Einstein equation
(\ref{eqn:bulkeq-tensor}) becomes
%
\begin{equation}
 l^2e^A\left[ e^{-3A}(e^{2A}F_{(T)}^{[1]})^{\cdot}\right]^{\cdot}
  -F_{(T)}^{[0]} = 0. 
\end{equation}
Hence, we obtain
%
\begin{equation}
  l^2e^{-3A}(e^{2A}F_{(T)}^{[1]})^{\cdot} = 
   \tilde{D}_1B_1(w) + \tilde{D}_2B_2(w) + \tilde{D}_3,
   \label{eqn:1st-integral-tensor}
\end{equation}
where $\tilde{D}_3$ is a constant and 
%
\begin{eqnarray}
 B_1(w) & = & 
  \int_0^wdw'e^{-3A(w')}\int_0^{w'}dw''e^{3A(w'')}, \nonumber\\
 B_2(w) & = & \int_0^wdw'e^{-3A(w')}. 
\end{eqnarray}
By integrating (\ref{eqn:1st-integral-tensor}) once more we obtain a 
solution of the bulk Einstein equation to this order. For the solution,
the junction condition (\ref{eqn:junction-tensor}) becomes
%
\begin{equation}
 \tilde{D}_1B_1(w_{\pm})+\tilde{D}_2B_2(w_{\pm})+\tilde{D}_3 = 
  \pm\kappa_5^2l^2e^{-2A_{\pm}-\alpha^{(0)}_{\pm}}
  \bar{\tau}^{[1]}_{\pm(T)}. 
\end{equation}
Hence, the constants $\tilde{D}_2$ and $\tilde{D}_3$ are uniquely
determined by $\bar{\tau}^{[0,1]}_{\pm(T)}$. In particular, 
%
\begin{equation}
 \tilde{D}_2 = \frac{\kappa_5^2l^2}{B_2(w_+)-B_2(w_-)}
  \left[e^{-2A_+-\alpha^{(0)}_+}\bar{\tau}_{+(T)}^{[1]}+
   e^{-2A_--\alpha^{(0)}_-}\bar{\tau}_{-(T)}^{[1]}\right]
  - \frac{B_1(w_+)-B_1(w_-)}{B_2(w_+)-B_2(w_-)}\tilde{D}_1. 
\end{equation}

Therefore, for small $|\mu|=|l^2\eta^{\mu\nu}k_{\mu}k_{\nu}|$, by using
the formula (\ref{eqn:bar-f}), the doubly gauge invariant variable
$\bar{f}_{\pm(T)}$ representing perturbations of the physical metric can
be given by 
%
\begin{eqnarray}
 e^{2A_{\pm}+\alpha_{\pm}^{(0)}}\bar{f}_{\pm(T)} & = & -\tilde{D}_1
   \frac{\int_{w_-}^{w_+}dw'e^{-3A(w')}
   \int_{w_{\pm}}^{w'}dw''e^{3A(w'')}}
   {\int^{w_+}_{w_-}dw'e^{-3A(w')}}  \nonumber\\
 & &  + \frac{\kappa_5^2l^2}{\int^{w_+}_{w_-}dw'e^{-3A(w')}}
  \left[e^{-2A_+-\alpha^{(0)}_+}\bar{\tau}_{+(T)}^{[1]}
   +e^{-2A_--\alpha^{(0)}_-}\bar{\tau}_{-(T)}^{[1]}\right] + O(\mu),
\end{eqnarray}
where $\tilde{D}_1$ is given by (\ref{eqn:C1-tauT0}) and
$\bar{\tau}_{\pm(T)}^{[0]}$ must satisfy the relation
(\ref{eqn:relation-tauT}).


\section{Comparison with four-dimensional Einstein gravity}
\label{sec:comparison}

In this section we compare gravity in the brane world scenario with
$4$-dimensional Einstein gravity. We show that $4$-dimensional
Einstein gravity is restored at low energies on our brane, provided that
matter fields on the hidden brane are not excited. This qualitative
result does not depend on which brane we are living. However,
quantitatively, the effective $4$-dimensional gravitational constants on
the two branes are different.

\subsection{Perturbations in four-dimensional Einstein gravity}
\label{subsec:4DEinstein}

For the purpose of comparison, we review linearized Einstein
gravity in $4$-dimensions in terms of gauge-invariant variables. We
consider general perturbations around the Minkowski spacetime. Namely,
we consider the metric
%
\begin{equation}
 ds_4^2 = (\bar{q}^{(0)}_{\mu\nu}+\delta \bar{q}_{\mu\nu})
  dy^{\mu}dy^{\mu}, \nonumber\\
\end{equation}
where
%
\begin{equation}
 \bar{q}_{\mu\nu}^{(0)}  =  \Omega^2\eta_{\mu\nu}, 
\end{equation}
$\Omega$ is a non-zero constant conformal factor , and 
%
\begin{equation}
 \delta \bar{q}_{\mu\nu} = 
   \bar{\sigma}_{(T)}T_{(T)\mu\nu} 
   + \bar{\sigma}_{(LT)}T_{(LT)\mu\nu}
   + \bar{\sigma}_{(LL)}T_{(LL)\mu\nu} 
   + \bar{\sigma}_{(Y)}T_{(Y)\mu\nu}.
\end{equation}
Here, the coefficients $\sigma_{(T)}$, $\sigma_{(LT)}$,
$\sigma_{(LL)}$, and $\sigma_{(Y)}$ are constants. 
Following the previous section, we can construct gauge-invariant
variables $\bar{f}_{(T)}$ and $\bar{f}$.  
%
\begin{eqnarray}
 \bar{f}_{(T)} & = & \bar{\sigma}_{(T)}, \nonumber\\
 \bar{f} & = & \bar{\sigma}_{(Y)} 
  + \frac{1}{2}\eta^{\mu\nu}k_{\mu}k_{\nu}\bar{\sigma}_{(LL)}. 
\end{eqnarray}

As for the stress energy tensor $\bar{S}_{\mu\nu}$, we consider it
as a first order quantity and expand it as follows. 
%
\begin{equation}
 \bar{S}_{\mu\nu} = 
  \bar{\tau}_{(T)}T_{(T)\mu\nu} + \bar{\tau}_{(LT)}T_{(LT)\mu\nu} 
   + \bar{\tau}_{(LL)}T_{(LL)\mu\nu} + \bar{\tau}_{(Y)}T_{(Y)\mu\nu}, 
\end{equation}
where the coefficients $\bar{\tau}_{(T,LT,LL,Y)}$ are constants. In the
Minkowski background these coefficients are gauge-invariant by
themselves.

We can expand the four-dimensional Einstein equation
$\bar{G}_{(4)\mu\nu}=8\pi G_N\bar{S}_{\mu\nu}$ up to first order 
in the perturbations, where $\bar{G}_{(4)\mu\nu}$ is the Einstein tensor
constructed from the metric $\bar{q}_{\mu\nu}$, and express it in terms
of the above gauge-invariant variables. 
The
Einstein equation is 
%
\begin{eqnarray}
 2\bar{\tau}_{(Y)} & = & 
  3\eta^{\mu\nu}k_{\mu}k_{\nu}\bar{\tau}_{(LL)}, \nonumber\\
 \bar{f} & = & -16\pi G_N\Omega^2\bar{\tau}_{(LL)}
  \label{eqn:einstein-scalar}
\end{eqnarray}
for scalar perturbations, 
%
\begin{equation}
 \bar{\tau}_{(LT)} = 0
\end{equation}
for vector perturbations, and 
%
\begin{equation}
 \bar{q}^{(0)\mu\nu}k_{\mu}k_{\nu}\bar{f}_{(T)} 
  = 16\pi G_N\bar{\tau}_{(T)}
  \label{eqn:einstein-tensor}
\end{equation}
for tensor perturbations.

Note that in such a gauge~\footnote{In (3) and (4) we adopted this gauge
choice.} that $\partial^{\mu}(\delta\bar{q}_{\mu\nu}
-\bar{q}^{(0)}_{\mu\nu}\bar{q}^{(0)\rho\sigma}\delta\bar{q}_{\rho\sigma}/2)
=0$, the variable $\bar{f}$ is given by
$\bar{f}=(2/3)\bar{\sigma}_{(Y)}$. Hence, in this gauge the second 
equation of (\ref{eqn:einstein-scalar}) becomes 
%
\begin{equation}
 \bar{q}^{(0)\mu\nu}k_{\mu}k_{\nu}\bar{\sigma}_{(Y)} 
  = -16\pi G_N\bar{\tau}_{(Y)}
  \label{eqn:einstein-scalar-fixed}
\end{equation}
It is easy to confirm by noting
$\Box=-\bar{q}^{(0)\mu\nu}k_{\mu}k_{\nu}$ that equations
(\ref{eqn:einstein-tensor}) and (\ref{eqn:einstein-scalar-fixed}) are
equivalent to (\ref{eqn:BD-linear-decomposed}) with $B=0$. Of course it
is also easy to derive Newton's law from equation (3) with $\Delta
K=B=0$~\cite{Wald}.

\subsection{Perturbations in the brane world scenario and restoration of
  Einstein gravity}

Here we shall apply the results of Sec.~\ref{sec:perturbations} to the
case that our $4$-dimensional universe is one of the two branes
$\Sigma_{\pm}$ and that there is no matter excitation on the other
brane.

First, let us suppose that our $4$-dimensional universe is $\Sigma_+$
and that there is no matter excitation on the other brane
$\Sigma_-$. For scalar perturbations, by setting
$\bar{\tau}_{-(Y)}=\bar{\tau}_{-(LL)}=0$, we obtain 
%
\begin{equation}
 \bar{f}_{+} = - \frac{\kappa_5^2e^{-2A_{+}-\alpha^{(0)}_{+}}}
  {\int^{w_+}_{w_-}dw'e^{-3A(w')}}
  e^{-2A_{+}-\alpha^{(0)}_{+}}\bar{\tau}_{+(LL)}
\end{equation}
for zero modes. This equation can be considered the lowest order
equation in an expansion in terms of
$\mu=l^2\eta^{\mu\nu}k_{\mu}k_{\nu}$, where $l$ is a characteristic
length scale of the model~\footnote{See eq.~(\ref{eqn:expansion-mu}) for
tensor perturbations and the footnote before it.}. Moreover, we have the
conservation equation 
%
\begin{equation}
 2\bar{\tau}_{+(Y)} = 
  3\eta^{\mu\nu}k_{\mu}k_{\nu}\bar{\tau}_{+(LL)}
\end{equation}
for all modes.

For vector perturbations, by setting
$\bar{\tau}_{-(LT)}=0$ we obtain 
%
\begin{equation}
 \bar{\tau}_{+(LT)} = 0
\end{equation}
for general $k_{\mu}$. Here, we have used the relation
(\ref{eqn:relation-vector}) for zero modes, or modes with
$\eta^{\mu\nu}k_{\mu}k_{\nu}= 0$. For tensor perturbations, by setting 
$\bar{\tau}_{-(T)}^{[0]}=\bar{\tau}_{-(T)}^{[1]}=0$, we obtain 
%
\begin{equation}
 \bar{q}^{(0)\mu\nu}_{+}k_{\mu}k_{\nu}\bar{f}_{+(T)} 
 = \frac{\kappa_5^2e^{-2A_+-\alpha^{(0)}_+}}{\int^{w_+}_{w_-}dw'e^{-3A(w')}}
 \bar{\tau}_{+(T)}
 \label{eqn:eq-on-brane-tensor}
\end{equation}
for small but non-zero $|\mu|=|l^2\eta^{\mu\nu}k_{\mu}k_{\nu}|$. Here,
we have used the relation (\ref{eqn:relation-tauT}) to show that
$\bar{\tau}_{+(T)}^{[0]}=0$. Actually,
eq.~(\ref{eqn:eq-on-brane-tensor}) remains correct even for
$\eta^{\mu\nu}k_{\mu}k_{\nu} \to 0$ since $D_2$ in (\ref{eqn:fT-zeromode})
is arbitrary and the relation (\ref{eqn:relation-tensor}) implies
$\bar{\tau}_{+(T)}=0$.

Next, let us suppose that our $4$-dimensional universe is $\Sigma_-$
and that there is no matter excitation on the other brane $\Sigma_+$. 
Repeating the above analysis, we can obtain the same result with $\pm$ 
replaced by $\mp$. For scalar perturbations, we have
%
\begin{eqnarray}
 2\bar{\tau}_{-(Y)} & = & 
  3\eta^{\mu\nu}k_{\mu}k_{\nu}\bar{\tau}_{-(LL)}, \nonumber\\
 \bar{f}_{-} & = & - \frac{\kappa_5^2e^{-2A_{-}-\alpha^{(0)}_{-}}}
  {\int^{w_+}_{w_-}dw'e^{-3A(w')}}
  e^{-2A_{-}-\alpha^{(0)}_{-}}\bar{\tau}_{-(LL)}
\end{eqnarray}
for zero modes.

For vector perturbations, we have
%
\begin{equation}
 \bar{\tau}_{-(LT)} = 0
\end{equation}
for general $k_{\mu}$. For tensor perturbations, we have
%
\begin{equation}
 \bar{q}^{(0)\mu\nu}_{-}k_{\mu}k_{\nu}\bar{f}_{-(T)} 
 = \frac{\kappa_5^2e^{-2A_--\alpha^{(0)}_-}}{\int^{w_+}_{w_-}dw'e^{-3A(w')}}
  \bar{\tau}_{-(T)}
\end{equation}
for $|\mu|$ zero or sufficiently small.

Therefore, by setting $\Omega^2=e^{-2A_{\pm}-\alpha^{(0)}_{\pm}}$, 
$4$-dimensional Einstein gravity is restored at low energies, provided
that there are no tachyonic modes, or positive $\mu$ modes, with
$\bar{\tau}_{\pm(T,LT,LL,Y)}=0$. The condition
$\bar{\tau}_{\pm(T,LL,Y)}=0$ for scalar and tensor perturbations 
corresponds to the boundary condition
%
\begin{eqnarray}
 \left.\tilde{B}_{\pm}\dot{F} 
  + e^{2A}\eta^{\mu\nu}k_{\mu}k_{\nu}\dot{\Psi}F\right|_{w=w_{\pm}} & = & 0, 
 \nonumber\\
 \left.(e^{2A}F_{(T)})^{\cdot}\right|_{w=w_{\pm}} & = & 0,
  \label{eqn:tau=0-bc}
\end{eqnarray}
where $\tilde{B}_{\pm}$ is given by (\ref{eqn:BandtildeB}). 
Hence, we need to show the absence of tachyonic modes given this
boundary condition. For this purpose we need to specify model potentials
and background solutions since it seems difficult to obtain analytic
solutions to the bulk equations (\ref{eqn:scalar-bulkeq}) and 
(\ref{eqn:bulkeq-tensor}) for a general positive
$\eta^{\mu}k_{\mu}k_{\nu}$, general potentials, and a general
background. On the other hand, we have already shown that there are no
tachyonic modes for vector perturbations (see
eq.~(\ref{eqn:vector-solution})). The effective $4$-dimensional Newton's
constants $G_{N\pm}$ on $\Sigma_{\pm}$ are given by 
%
\begin{equation}
 16\pi G_{N\pm} = 
  \frac{\kappa_5^2e^{-2A_{\pm}-\alpha^{(0)}_{\pm}}}
  {\int^{w_+}_{w_-}dw'e^{-3A(w')}}. 
  \label{eqn:Newton-const}
\end{equation}
Since $w_-<w_+$, $G_{\pm N}$ is positive on both branes~\footnote{If we
had choose the convention $w_->w_+$ then the sign in
(\ref{eqn:unit-normal}) would be reversed and $G_{\pm N}$ would still be
positive, as should be.}. It is worth while mentioning that the
following naive expectation holds. 
%
\begin{equation}
 \frac{G_{N+}}{G_{N-}} = 
  \frac{e^{-2A_+-\alpha^{(0)}_+}}{e^{-2A_--\alpha^{(0)}_-}}
  = \frac{\bar{q}^{(0)}_{+\mu\nu}}{\bar{q}^{(0)}_{-\mu\nu}},
  \label{eqn:warp-factor}
\end{equation}
where $\bar{q}^{(0)}_{\pm\mu\nu}$ are the unperturbed physical metrics
on $\Sigma^{(0)}_{\pm}$, respectively.


\section{Summary }
\label{sec:summary}

We have investigated gravity in brane worlds. In particular, we
considered a general model on a $Z_2$-orbifold with a bulk scalar field
and investigated gauge-invariant perturbations around a background with
two parallel Minkowski branes. By using the doubly covariant formalism
of perturbation of Israel's junction conditions~\cite{Israel} developed
in ref.~\cite{Mukohyama2000b}, we have obtained equations governing the
low energy dynamics of doubly gauge invariant perturbations on the
branes. These doubly gauge invariant perturbations include scalar,
vector, and tensor perturbations. After that, we have shown that
$4$-dimensional Einstein gravity is restored on both branes at low 
energies, provided that combinations of the bulk and the brane
potentials  $\tilde{B}_{\pm}$ given by (\ref{eqn:BandtildeB}) are not
zero at the branes and that there are no tachyonic modes for scalar
perturbations with the boundary condition (\ref{eqn:tau=0-bc}). 
It is easy to see that the condition $\tilde{B}_{\pm}\ne 0$ is
equivalent to the absence of a scalar zero mode with vanishing matter on
the branes. Actually, the would-be zero mode with vanishing matter is
prohibited by the boundary condition (\ref{eqn:C1-C2-bc}) with
$\bar{\tau}_{\pm(LL)}=0$ if and only if $\tilde{B}_{\pm}=0$.

The effective $4$-dimensional Newton's constants $G_{N\pm}$ on 
$\Sigma_{\pm}$, respectively, are given by (\ref{eqn:Newton-const}) and
the naive  expectation (\ref{eqn:warp-factor}) holds.  Thus we have
confirmed and extended the result of \cite{Tanaka-Montes} that pure
Einstein gravity is restored for the braneworld scenario with a single
bulk scalar at low energies.

The condition $\tilde{B}_{\pm}\ne 0$ is equivalent to the condition that
the inter-brane distance (radion) is stabilized (to be precise, fixed)
and that the background bulk scalar field
has non-zero derivatives  near the branes. To see this it is enough to
differentiate the background scalar-field matching condition (the second
of (\ref{eqn:background-junction})) with respect to
$w_{\pm}$~\cite{Mukohyama2001c}. In fact, the radion stabilization can
be stated as
%
\begin{eqnarray}
 0 \neq e^{A_{\pm}}\frac{\partial}{\partial w_{\pm}}
  \left[\left.e^A\dot{\Psi}^{(0)}\right|_{w=w_{\pm}} 
  \pm\frac{1}{2}V'_{\pm}(\Psi^{(0)}_{\pm})\right]
   = {U'_{\pm}}^{(0)}
     +\frac{\kappa_5^2}{3}V_{\pm}^{(0)}{V'_{\pm}}^{(0)}
                 -\frac{1}{4}{V'_{\pm}}^{(0)}{V''_{\pm}}^{(0)}
		 \label{eqn:stabilization}
\end{eqnarray}
since the following equation is always satisfied because of the
background equation. 
%
\begin{eqnarray}
 \frac{\partial}{\partial w_{\pm}}
 \left[ \left.e^A\dot{A}\right|_{w=w_{\pm}} 
  \pm\frac{1}{6}\kappa_5^2V_{\pm}(\Psi^{(0)}_{\pm})\right]=0.
\end{eqnarray}
Essentially, the condition (\ref{eqn:stabilization}) says that the
position of the brane cannot be changed without changing the bulk
solution. Hence, actually the condition $\tilde{B}_{\pm}\ne 0$ is 
equivalent to the condition that $V^{'(0)}_{\pm}\ne 0$ and that the
inter-brane distance is fixed.
Since $V^{'(0)}_{\pm}$ is related to
$\dot{\Psi}^{(0)}$ by (\ref{eqn:background-junction}) on branes, finally
the condition $\tilde{B}_{\pm}\ne 0$ is equivalent to the condition that
the background bulk scalar field is changing near the branes and that
the radion is stabilized.

One might expect that it is sufficient but not necessary to fix
positions of both branes in order to fix the distance between two
branes. In fact, it is necessary to fix positions of both branes as in 
(\ref{eqn:stabilization}). To see this, let us assume that the
position of one brane can be changed without changing the bulk solution
at all. (The position of the other brane can be either fixed or
unfixed.) In this case it is impossible for the position of the unfixed 
brane to affect the position of the other brane since the only possible
medium of the communication between the two branes would be the bulk
but the bulk solution is independent of the position of the unfixed
brane by  assumption. Thus, the distance between the branes is not fixed
in this  case. In other words, the matching condition for the background
with the $4$-dimensional Poincare symmetry is local.

Although to derive Newton's gravitational couplings, we have used only
matter source at the brane $\Sigma_+$ (visible brane), it is instructive
to write down the most generic weak gravity equations taking into
account the contribution from the hidden brane, which enters the final
results in a symmetric manner, see the last equations in subsections 
\ref{subsec:scalar} and \ref{subsec:tensor}. Suppose energy-momentum
tensor at the visible brane is $T^+_{\mu\nu}$ and that on the hidden
brane  $\Sigma_-$ is $T^-_{\mu\nu}$. Let us again, as in Sec. II, use
linearized metric perturbations $h_{\mu\nu}$ at the visible
brane. Similar to (\ref{eqn:BD-linear-decomposed}), in most general case
we have 
\begin{eqnarray}
 \Box \bar{h}^{TT}_{\mu\nu} & = & -16 \pi G^+_N  T^{+TT}_{\mu\nu}
  -16 \pi G^-_N  T^{-TT}_{\mu\nu} \ ,
  \nonumber\\
 \Box h & = & {16 \pi G^+_N}(1-2B^+)T^+ +  {16 \pi G^-_N}(1-2B^-)T^-
 \ .\label{eqn:BD-linear-decomposed1}
\end{eqnarray}
Therefore, in principle, we need four gravitational couplings,
two for tensor modes and two for scalar modes,
to describe the linearized gravity at the visible brane.
We also learned that each of them are constructed from the
fundamental 5d gravitational constant, $\kappa_5$,
and  combinations of the warp factor and ``effective''
extra dimensional distance $l \sim 
  {\int^{w_+}_{w_-}dw'e^{-3A(w')}}$. It turns out that
the BD parameter, which characterizes the
gravitational coupling for the scalar mode, depends on the bulk scalar sector.
 In terms of coefficient $B$ of Sec. II,
$B^{\pm}$ is none-zero for  the case without bulk scalar,
$B^{\pm}$=0 for the case of a single scalar field.
We conjecture, based on  our derivation with a single scalar,
that in general case of more than one bulk scalar,
$B^{\pm}$ again will be non-zero.
It will be interesting to calculate it explicitly
for next simplest case of two bulk scalars.

Equation (\ref{eqn:BD-linear-decomposed1}) may have interesting insights
from point of view of superstring (Horava-Witten) phenomenology. Indeed,
it says that there are, in general, four different Planckian masses! It
also teaches us that derivation of 4d effective theory by means of 
$\int dw$ integration will wash out the subtleties of the actual
gravitational dynamics. Finally, from point of view of gravity and its
application to astronomy, it gives us rather unusual theory with four
gravitational couplings.

\begin{acknowledgments}
 We are grateful to Gary Felder, Andrei Frolov, Andrei Linde, Johannes
 Martin and Takahiro Tanaka for useful discussions. One of the author
 (S.M.) would like to thank Werner Israel for continuing
 encouragement. S.M.'s work is supported by a JSPS Postdoctoral
 Fellowship for Research Abroad, L.K. was supported by NSERC and CIAR. 
\end{acknowledgments}


\appendix


\section{Harmonics in Minkowski spacetime}
\label{app:harmonics}

In this appendix we give definitions of scalar, vector and tensor
harmonics in an $n$-dimensional Minkowski spacetime. Throughout this 
appendix, $n$-dimensional coordinates are $x^{\mu}$
($\mu=0,1,\cdots,n-1$), $\eta_{\mu\nu}$ is the Minkowski metric, and all
indices are raised and lowered by the Minkowski metric and its inverse
$\eta^{\mu\nu}$.

\subsection{Scalar harmonics}

The scalar harmonics are given by 
%
\begin{equation}
 Y = exp(-ik_{\rho}x^{\rho}),
\end{equation}
by which any function $f$ can be expanded as 
%
\begin{equation}
 f = \int dk\ c Y,
\end{equation}
where $c$ is a constant depending on $k$. Hereafter, $k$ and $dk$ are
abbreviations of $\{k^{\mu}\}$ ($\mu=0,1,\cdots,n-1$) and 
$\prod_{\mu=0}^{n-1}dk^{\mu}$, respectively. We omit $k$ in most cases. 

\subsection{Vector harmonics}

In general, any vector field $v_{\mu}$ can be decomposed as
%
\begin{equation}
 v_{\mu}=v_{(T)\mu}+\partial_{\mu} f ,
\end{equation}
where $f$ is a function and $v_{(T)\mu}$ is a transverse vector field:
%
\begin{equation}
 \partial^{\mu}v_{(T)\mu}=0 .
\end{equation}

Thus, the vector field $v_{\mu}$ can be expanded by using the scalar 
harmonics $Y$ and transverse vector harmonics $V_{(T)\mu}$ as 
%
\begin{equation}
 v_{\mu} = \int dk
  \left[c_{(T)}V_{(T)\mu}+c_{(L)}\partial_{\mu} Y\right].
	\label{eqn:dY+V}
\end{equation}
Here, $c_{(T)}$ and $c_{(L)}$ are constants depending on $k$, and the
transverse vector harmonics $V_{(T)\mu}$ are given by 
%
\begin{equation}
 V_{(T)\mu} = u_{\mu}\exp(-ik_{\rho}x^{\rho}),
\end{equation}
where the constant vector $u_{\mu}$ satisfies the following condition. 
%
\begin{equation}
 k^{\mu}u_{\mu}=0
\end{equation}
for $k^{\mu}k_{\mu}\ne 0$, and 
%
\begin{eqnarray}
 k^{\mu}u_{\mu} & = & 0, \nonumber\\
 \tau^{\mu}u_{\mu} & = & 0
  \label{eqn:u-for-k^2=0}
\end{eqnarray}
for non-vanishing $k_{\mu}$ satisfying $k^{\mu}k_{\mu}=0$, where
$\tau^{\mu}$ is an arbitrary constant timelike vector. 
For $k_{\mu}=0$,
the constant vector $u^{\mu}$ does not need to satisfy any of the above
conditions. For the special case $k^{\mu}k_{\mu}=0$, the second
condition in (\ref{eqn:u-for-k^2=0}) can be imposed by redefinition of
$c_{(L)}$. Actually this condition is necessary to eliminate
redundancy. Note that the number of independent vectors satisfying the
above condition is $n-1$ for $k^{\mu}k_{\mu}\ne 0$ and $n-2$ for
$k^{\mu}k_{\mu}=0$ and that these numbers are equal to the numbers of
physical degrees of freedom for massive and massless spin-$1$ fields in
$n$-dimensions, respectively.

Because of the expansion (\ref{eqn:dY+V}), it is convenient to define
longitudinal vector harmonics $V_{(L)\mu}$ by
%
\begin{equation}
 V_{(L)\mu} \equiv \partial_{\mu} Y = -ik_{\mu}Y. 
\end{equation}

\subsection{Tensor harmonics}

In general, a symmetric second-rank tensor field $t_{\mu\nu}$ can be
decomposed as
%
\begin{equation}
 t_{\mu\nu}=t_{(T)\mu\nu} + \partial_{\mu}v_{\nu}+\partial_{\nu}v_{\mu} 
  + f\eta_{\mu\nu},
\end{equation}
where $f$ is a function, $v_{\mu}$ is a vector field and $t_{(T)\mu\nu}$
is a transverse traceless symmetric tensor field:
%
\begin{eqnarray}
 t_{(T)\mu}^\mu & = & 0,\nonumber\\
 \partial^{\mu} t_{(T)\mu\nu} & = &0.
	\label{eqn:trasverse-traceless}
\end{eqnarray}

Thus, the tensor field $t_{\mu\nu}$ can be expanded by using the scalar
harmonics $Y$, the vector harmonics $V_{(T)}$ and $V_{(L)}$, and
transverse traceless tensor harmonics $T_{(T)}$ as 
%
\begin{eqnarray}
 t_{\mu\nu} & = & \int dk\left[
	c_{(T)}T_{(T)\mu\nu}+c_{(LT)}
	(\partial_{\mu}V_{(T)\nu}+\partial_{\nu}V_{(T)\mu})\right.
	\nonumber\\
 & &	\left.
	 + c_{(LL)}(\partial_{\mu}V_{(L)\nu}+\partial_{\nu}V_{(L)\mu})
	+ \tilde{c}_{(Y)}Y\eta_{\mu\nu}\right].
	\label{eqn:dV+T}
\end{eqnarray}
Here, $c_{(T)}$, $c_{(LT)}$, $c_{(LL)}$, and $\tilde{c}_{(Y)}$ are
constants depending on $k$, and the transverse traceless tensor
harmonics $T_{(T)}$ are given by 
%
\begin{equation}
 T_{(T)\mu\nu} = s_{\mu\nu}\exp(-ik_{\rho}x^{\rho}),
\end{equation}
where the constant symmetric second-rank tensor $s_{\mu\nu}$ satisfies
the following condition. 
%
\begin{eqnarray}
 k^{\mu}s_{\mu\nu} & = & 0, \nonumber\\
 s^{\mu}_{\mu} & = & 0
\end{eqnarray}
for $k^{\mu}k_{\mu}\ne 0$, and 
%
\begin{eqnarray}
 k^{\mu}s_{\mu\nu} & = & 0, \nonumber\\
 s^{\mu}_{\mu} & = & 0, \nonumber\\
 \tau^{\mu}s_{\mu\nu} & = & 0
  \label{eqn:s-for-k^2=0}
\end{eqnarray}
for non-vanishing $k_{\mu}$ satisfying $k^{\mu}k_{\mu}=0$, where
$\tau^{\mu}$ is an arbitrary constant timelike vector. 
For $k_{\mu}=0$, the constant tensor $s_{\mu\nu}$ does not need to
satisfy any of the above conditions. For the special case
$k^{\mu}k_{\mu}=0$, the last condition in (\ref{eqn:s-for-k^2=0}) can be
imposed by redefinition of $c_{(LT)}$, $c_{(LL)}$ and
$\tilde{c}_{(Y)}$. Actually this condition is 
necessary to eliminate redundancy. Note that the number of independent
symmetric second-rank tensors satisfying the above conditions is
$(n+1)(n-2)/2$ for $k^{\mu}k_{\mu}\ne 0$ and $n(n-3)/2$ for 
$k^{\mu}k_{\mu}=0$ and that these numbers are equal to numbers of
physical degrees of freedom for massive and massless spin-$2$ fields in
$n$-dimensions, respectively.

Because of the expansion (\ref{eqn:dV+T}), it is convenient to define
tensor harmonics $T_{(LT)}$, $T_{(LL)}$, and $T_{(Y)}$ by 
%
\begin{eqnarray}
 T_{(LT)\mu\nu} & \equiv & \partial_{\mu}V_{(T)\nu}
  +\partial_{\nu}V_{(T)\mu}, \nonumber\\
 & = & -i(u_{\mu}k_{\nu}+u_{\nu}k_{\mu})Y, \nonumber\\
 T_{(LL)\mu\nu} & \equiv & \partial_{\mu}V_{(L)\nu}
  +\partial_{\nu}V_{(L)\mu}
	-\frac{2}{n}\eta_{\mu\nu}\partial^{\rho}V_{(L)\rho} \nonumber\\
 & = & \left(-2k_{\mu}k_{\nu}+\frac{2}{n}k^{\rho}k_{\rho}
	\eta_{\mu\nu}\right)Y,  \nonumber\\ 
 T_{(Y)\mu\nu} & \equiv & \eta_{\mu\nu}Y. 
\end{eqnarray}


\end{document}